\documentclass[prb,twocolumn,nofootinbib,citeautoscript,10pt,longbibliography,notitlepage]{revtex4-2}
\usepackage{graphicx}
\usepackage{dcolumn}
\usepackage{bm}
\usepackage{amsmath}
\usepackage{tabularx,graphicx}
\usepackage{epstopdf}
\usepackage{latexsym}
\usepackage{amssymb}
\usepackage{amsmath}
\usepackage{color,colortbl,array}
\usepackage{psfrag}
\usepackage{bbm}
\usepackage{bm}
\usepackage{titlesec}
\usepackage{dsfont}
\usepackage{feynmp}
\usepackage{slashed}
\usepackage{multirow}
\usepackage[tight]{subfigure}

\usepackage[papersize={8.5in,11in}]{geometry}

\usepackage{color}
\definecolor{darkblue}{rgb}{0.,0.,0.4}
\definecolor{darkred}{rgb}{0.5,0.,0.}
\definecolor{BlueViolet}{RGB}{138,43,226}
\definecolor{SkyBlue}{RGB}{30,144,255}
\definecolor{DarkGreen}{RGB}{0,100,0}
\usepackage[pdftex,colorlinks=true,linkcolor=darkblue,citecolor=blue,urlcolor=darkred]{hyperref}

\usepackage{physics}

\def\avg#1{\left\langle#1\right\rangle}

\def\Re{{\rm Re}}

\def\nn{\nonumber}

\def\=>{\Rightarrow}
\def\>{\rightarrow}

\def\eye2{\mathbb{I}}

\def \nn{\nonumber \\}

\begin{document}

\title{Enhanced eigenvector sensitivity and algebraic classification of sublattice-symmetric exceptional points}

\author{Kang Yang}
\affiliation
{Department of Physics, Stockholm University, AlbaNova University Center, 106 91 Stockholm, Sweden}
\affiliation{Dahlem Center for Complex Quantum Systems and Fachbereich Physik, Freie Universit\"{a}t Berlin, 14195 Berlin, Germany}
\author{Ipsita Mandal}
\affiliation{Institute of Nuclear Physics, Polish Academy of Sciences, 31-342 Krak\'{o}w, Poland}

\begin{abstract}
Exceptional points (EPs) are degeneracy of non-Hermitian Hamiltonians, at which the eigenvalues, along with their eigenvectors, coalesce. Their orders are given by the Jordan decomposition. Here, we focus on higher-order EPs arising in fermionic systems with a sublattice symmetry, which restricts the eigenvalues of the Hamitlonian to appear in pairs of $\lbrace E,  -E\rbrace $.
Thus, a naive prediction might lead to only even-order EPs at zero energy. However, we show that odd-order EPs can exist and exhibit enhanced sensitivity in the behaviour of eigenvector-coalescence in their neighbourhood, depending on how we approach the degenerate point. The odd-order EPs can be understood as a mixture of higher- and lower-valued even-order EPs. Such an anomalous behaviour is related to the irregular topology of the EPs as the subspace of the Hamiltonians in question, which is a unique feature of the Jordan blocks. The enhanced eigenvector sensitivity can be described by observing how the quantum distance to the target eigenvector converges to zero. In order to capture the eigenvector-coalescence, we provide an algebraic method to describe the conditions for the existence of these EPs. This complements previous studies based on resultants and discriminants, and unveils heretofore unexplored structures of higher-order exceptional degeneracy.
\end{abstract}

\maketitle

\tableofcontents

\section{Introduction}

Exceptional degeneracy is a phenomenon where the eigenvalues of a matrix cross each other and their eigenvectors collapse simultaneously, losing the linear independence \cite{Berry2004,Heiss_2012,PhysRevX.6.021007,ep-optics,ozdemir2019parity}. The simplest example is when two eigenvalues and their corresponding eigenvectors coalesce, leading to an exceptional point (EP) of second order. Such singularities can arise in the context of a great variety of physical problems, such as dissipative processes captured by non-Hermitian Hamiltonians \cite{plenio1998quantum,Daley2014,PhysRevX.9.041015,ding2022non,PhysRevLett.121.026403,PhysRevLett.125.227204,kang-emil,PhysRevB.99.201107,PhysRevB.100.245205,PhysRevX.8.031079,rev-emil,yang2022exceptional,PhysRevB.101.085122,PhysRevB.104.L121109,PhysRevA.103.L020201,PhysRevLett.126.083604,PhysRevLett.127.106601,PhysRevLett.128.160401}, and topological phase transitions in chiral Hamiltonians \cite{ips-tewari,ips-ep-epl}.
Their singular behaviour manifests itself in enhanced sensitivity, and thus has potential applications in detection and sensors \cite{PhysRevLett.112.203901,hodaei2017enhanced,chen2017exceptional,PhysRevA.98.023805,wang2020petermann,park2020symmetry,PhysRevLett.125.180403}.

An $n^{\rm{th}}$-order exceptional point (EP$_n$) \cite{demange2011signatures,jing2017high,PhysRevLett.117.107402,PhysRevA.101.033820,ips-prl,PhysRevLett.127.186602,PhysRevA.104.063508,sayyad2022realizing} appears when the Jordan decomposition of the matrix contains an $n$-dimensional (with $n>1$) Jordan block $J_n(E)$ along its diagonal, at the eigenvalue $E$.
Near an EP$_2$, the dispersion varies as a square root, viz.,  $\delta E\sim \sqrt{|\delta \mathbf q|}$, where $|\delta \mathbf q|$ characterizes the deviation from the EP in the momentum space spanned by the vector $\mathbf q$. 
The derivative of the dispersion diverges at the EP, implying that the change in eigenvalue becomes more and more sensitive as we approach the EP. Such a sensitivity is further enhanced at a higher-order EP$_n$ ($n>2$), because now an $n^{\rm{th}}$-order root sensitivity (i.e., $\delta E\sim |\delta \mathbf q|^{1/n}$) can appear in the vicinity of the EP$_n$ for generic situations \cite{ips-prl,PhysRevLett.127.186602,sayyad2022realizing}. The eigenvalue overlap at higher-order EPs can be captured by equations involving discriminants \cite{sayyad2022realizing} or resultants \cite{PhysRevLett.127.186602}. However, another important and unique property of an EP, namely the coalescence of eigenstates, remains elusive under this approach. Moreover, the space spanned by the exceptional degeneracy is not a closed subspace of the parameter space of the corresponding matrix \cite{PhysRevA.102.032216}. In fact, this space has a finer topological structure beyond the solutions captured by continuous functions (such as the discriminants and resultants) of the matrix.

\begin{figure*}[]
	\centering
\includegraphics[width=0.9\textwidth]{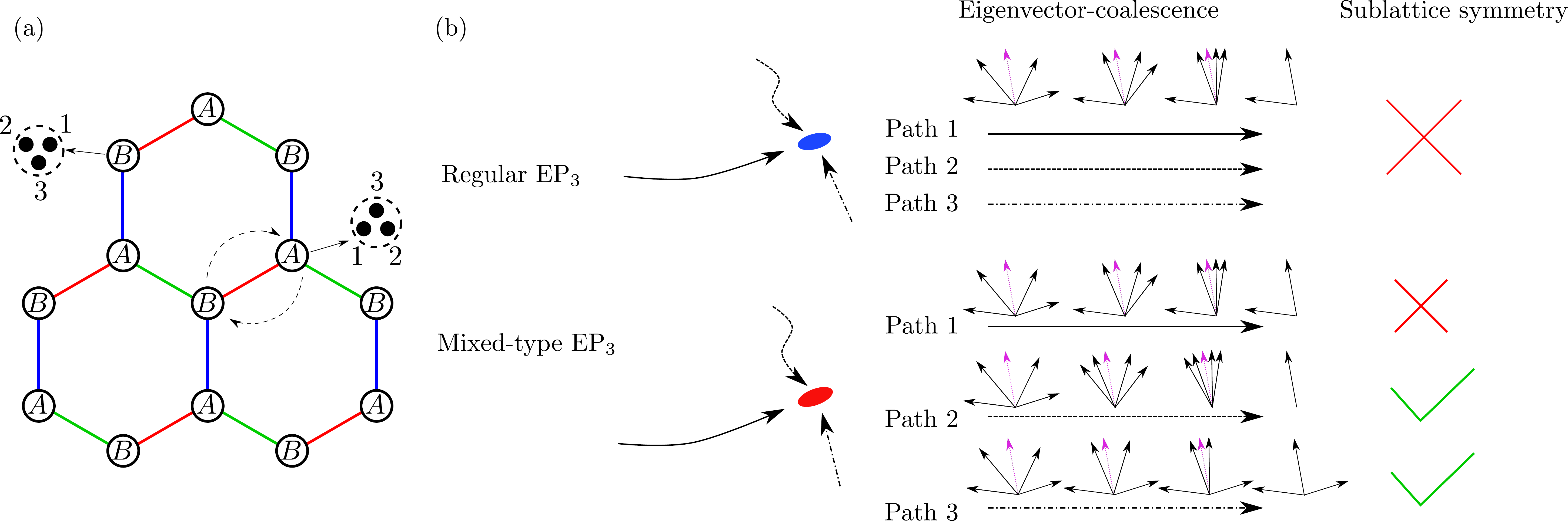}
\caption{\label{fighoney}
(a) Decorated honeycomb lattice model of fermions with $N=3$ flavours \cite{yao-lee}, also dubbed as the ``Yao-Lee'' model (see Appendix~\ref{secsupp4}). The system has a sublattice symmetry when only nearest-neighbour hoppings are included in the Hamiltonian. The fermions are labelled by their sublattice indices $A$ and $B$, together with their flavour index $\alpha \in \lbrace 1,2, 3 \rbrace$ on each sublattice site. (b) The coalescence of eigenvectors for a four-band model near a regular EP$_3$ (blue oval disc) and a mixed-type EP$_3$ (red oval disc). Near the regular EP$_3$, three eigenvectors out of the four are collapsing to a single eigenvector at the EP. Near the mixed-type EP$_3$, how the eigenvectors coalesce strongly depends on the path chosen to approach the EP. There can be two-fold, three-fold, and four-fold eigenvector-coalescence for the three different paths indicated by the dash-dotted, solid, and dashed lines, respectively. When sublattice symmetry is imposed, the three-fold eigenvector-coalescence is forbidden.
}
\end{figure*}

In this paper, we use an algebraic method to classify the higher-order EPs according to their eigenvector-coalescence. We focus on the nature of the higher-order EPs that can appear in two-dimensional (2d) systems in the presence of a sublattice symmetry [cf. Fig.~\ref{fighoney}(a)], and determine how their eigenstates collapse. The main results are summarized in Fig.~\ref{fighoney}(b) and Table~\ref{tb_EPSpace}. Remarkably, according to our classification, all EP$_n$'s can be categorized into two types. A regular EP$_n$ exhibits a typical $n$-fold eigenvector-coalescence, while a mixed-type EP$_n$ can exhibit different eigenvector-coalescence depending on how our Hamiltonian is approaching it in the parameter space.

The model is implemented by considering $N$ flavours of fermions, living on a bipartite lattice, whose creation operators are given by ${c^\alpha_1}^{\dagger}$ and ${c^\alpha_2}^{\dagger}$($\alpha \in [1,N]$). The degrees of freedom for the two sublattices have been distinguished by the subscripts $1$ and $2$. The sublattice symmetry ensures that the Hamiltonian $H$ obeys $P \, H \,P=-H$ \cite{PhysRevB.78.195125,PhysRevB.93.085101}, with the operator $P$ acting as $c^\alpha_1 \xrightarrow{P} c^\alpha_1 $ and $c^\alpha_2 \xrightarrow{P} -c^\alpha_2$. This is a very natural condition when the Hamiltonian contains only hoppings from sublattice $1$ to sublattice $2$. Examples of such Hamiltonians include solvable spin liquid models, such as the Kitaev spin liquid \cite{kitaev} (corresponding to $N=1$), and the Yao-Lee $SU(2)$ spin liquid \cite{yao-lee} (corresponding to $N=3$). In Hermitian systems, the sublattice symmetry can be viewed as the product of time-reversal transformation and particle-hole transformation of fermions, which translates to a chiral symmetry \cite{PhysRevB.78.195125}. In the momentum space, a generic non-Hermitian Hamiltonian with the sublattice symmetry can be brought to the block off-diagonal form:
\begin{align}
H(\mathbf q) =\begin{pmatrix}
0 & i\,\mathbf{B}(\mathbf{q}) \\
-i\,\mathbf{B}'(\mathbf{q}) & 0 \\
\end{pmatrix},
\label{eq_NHYL}
\end{align}
where $\mathbf{B} $ and $\mathbf{B}'$ are $N\times N$ matrices.
In order to demonstrate our results in closed analytical forms, we will focus on the $N=2$ case, where the system can be described by $4 \times  4$ matrices.

We will characterize our EPs based on the nilpotency of Jordan blocks in the generalized eigenspace. 
To explain the terminologies, let us consider the example of an EP$_3$. Near an EP$_3$, we have a three-dimensional Jordan block, and the Hamiltonian can be expressed as
\begin{align}
V \,H(\mathbf q_\ast) \,V^{-1}&=\textrm{diag}
\{J_3(E_1), \,E_2, \,E_3, \,\dots\}\nonumber\\ &=
\begin{pmatrix}
E_1 & 1 & 0 & 0 &0 & \dots \\
0 & E_1 & 1 & 0 &0 &\dots\\ 
0 & 0 & E_1 & 0 &0 &\dots\\
0 & 0 & 0 & E_2 &0&\dots\\
0 & 0 & 0 & 0 & E_3 &\dots\\ 
\vdots & \vdots & \vdots &0 & 0 & \ddots\\ 
\end{pmatrix},
\end{align}
where $E_1$ is a three-fold degenerate eigenvalue with only one linearly independent eigenvector proportional to $e_1=V \,(1, \,0,  \,\dots)^T$. The generalized eigenspace $\mathcal L_{E_1}$ of $E_1$ includes two other vectors, viz.,  $e_2=V\,(0,\,1, \,0  , \,\dots)^T$, and $ e_3=V\,(0,\, 0, \,1, \,0, \, \dots)^T$, such that $(H-E_1)$ is nilpotent in $\mathcal L_{E_1}$. In other words, $(H-E_1) \,e_3=e_2$, $(H-E_1) \,e_2=e_1 $, and $(H-E_1)\, e_1=0$. Intuitively, this EP$_3$ is interpreted as the singular point where the three eigenvectors of $E_1$ collapse into one. According to the Jordan decomposition, we denote the point $\mathbf q =\mathbf q_\ast$ as a {\it simple} EP$_3$, if all the Jordan blocks belonging to the other eigenvalues $E_2, \,E_3, \,\dots$ are trivial (i.e., one-dimensional). If the Hamiltonian has more than one eigenvalue whose Jordan block is nontrivial (i.e., has dimension greater than unity), we denote the point $\mathbf q = \mathbf q_\ast$ as a {\it compound} EP.

The paper is organized as follows. In Sec.~\ref{sc_smEP}, we discuss the sublattice symmetry and the nature of the EPs, which is the main result of this paper. Sec.~\ref{sc_eq2d} focusses on the properties of various types of EPs and the analytical solutions of eigenvectors in their neighbourhoods. In Sec.~\ref{sc_todis}, we use a quantum distance to characterize the eigenvector folding near an EP, and explain the enhanced eigenvector sensitivity in terms of the unique subspace topology for non-Hermitian matrices. Sec.~\ref{secgen} deals with some explicit realizations of the systems discussed, and also touches upon the predictions for generic $N$-values. We conclude with a summary and outlook in Sec.~\ref{secsum}. Appendices~\ref{secsupp1}--\ref{secsupp4} show the details of the mathematical derivations of various results mentioned in the main text.


\section{Sublattice symmetry and the EP parameter space}
\label{sc_smEP}

The sublattice symmetry makes the characteristic polynomial of the Hamiltonian even in the eigenvalue $E$, as captured by the relation $\det(E-H)=\det[P\, (E-H)\, P]=\det(E+H) $, where we use the fact that the dimension of $H$ is even. The eigenvalues of $H$ thus always come in pairs of $\lbrace E,-E \rbrace $.  
A natural choice of basis under the sublattice symmetry is to group the upper and lower components of the eigenstates as $\psi$ and $\chi$, respectively. With this choice, the eigenvalue problem is reduced to the following equations:
\begin{align}
    E^2 \,\psi(\mathbf q)&=\mathbf B(\mathbf q) \cdot \mathbf B'(\mathbf q)\,
  \psi(\mathbf q),\, 
 E \,  \chi(\mathbf q)= -i \,\mathbf B'(\mathbf q) \,\psi(\mathbf q)\,,\nonumber\\
 E^2 \, \chi(\mathbf q)&=\mathbf B'(\mathbf q)  \cdot  \mathbf B(\mathbf q)\,\chi(\mathbf q)\,,\,  
 E \,\psi(\mathbf q)= i\, \mathbf B(\mathbf q)\,\chi(\mathbf q)\, .
\label{eq_ofdH}
\end{align}
The above indicates that if $( \psi^T , \, \chi^T )^T$ is an eigenvector for the eigenvalue $E$, $(  \psi^T  , \,- \chi^T  )^T$ is an eigenvector for $-E$. Besides eigenvalues, the sublattice symmetry also imposes the constraint that nondegenerate eigenvectors should appear in pairs.

The above pairing relation for eigenvectors at $E$ and $-E$  also applies to generalized eigenvectors, which include other linearly independent vectors in the generalized eigenspaces $\mathcal L_E$ and $\mathcal L_{-E}$, apart from the eigenvectors. If we take two generalized eigenvectors
$(\psi_1, \,\chi_1)$ and $(\psi_2, \,\chi_2)$, corresponding to an eigenvalue $E$ having a nontrivial Jordan block, then $(H-E)\,( \psi^T _2, \chi^T _2)^T=( \psi^T _1, \, \chi^T _1)^T$ is also in the generalized eigenspace of $E$. By applying $P$ to this equation, one can verify that $(H+E)\,(- \psi^T _2, \, \chi^T _2)^T=(  \psi^T _1,\,- \chi^T _1)^T$. Therefore, if $\lbrace ( \psi^T _1, \chi^T _1)^T$, $( \psi^T _2, \chi^T _2)^T$, $ ( \psi^T _3, \, \chi^T _3)^T, \,\dots \rbrace $ generate the generalized eigenspace $\mathcal L_E$, the eigenvectors $ \lbrace ( \psi^T _1, \,-  \chi^T _1)^T$, $(- \psi^T _2, \, \chi^T _2)^T$, $( \psi^T _3,\,-  \chi^T _3)^T, \,\dots \rbrace$ generate the generalized eigenspace $\mathcal L_{-E}$. 

According to the above analysis, the degeneracy of the system should be distinguished depending on whether it involves a zero or nonzero eigenvalue $E$ as follows:
\\\textbf{(1)} If all the eigenvalues are nonzero (i.e., $E\ne 0$), the lower component is linearly related to the upper component as $\chi=-i\,\mathbf B'(\mathbf q)\, \psi(\mathbf q)/E$. The problem is then entirely determined by the $2\times 2$ matrix $\mathbf B(\mathbf q)\cdot \mathbf B'(\mathbf q)$. If $E$ is an eigenvalue where two eigenvectors coalesce at the momentum $\mathbf q = \mathbf q_\ast$, then $-E$ shows an identical behaviour. Hence, the exceptional degeneracy for $E\ne 0$ must be a compound EP, always appearing as a doublet of EP$_2$'s.
\\\textbf{(2)} If $E=0$ is an eigenvalue with algebraic multiplicity $l$, the corresponding eigenvector is obtained from the kernels of the two matrices, i.e., those $\psi$ and $\chi$ which satisfy $\mathbf B(\mathbf q)\, \chi=0$ and $\mathbf B'(\mathbf q)\,\psi=0$. The eigenvectors are given by $(\psi^T,0)^T$ and $(0,\chi^T)^T$. Assuming that the numbers of solutions to the two equations are $\dim(\ker\mathbf B)=m$ and $\dim(\ker\mathbf B')=n$, respectively, we can construct $(m+n)$ distinct eigenvectors. Hence, the order of the EP can range from $2$ to $(l+1-m-n)$.

From the two possible cases, we find that the $E=0$ situation gives us the richest EP structure, and hence, this will be the focus of the rest of this paper. Denoting the eigenvalues of $\mathbf B(\mathbf q)\cdot \mathbf B'(\mathbf q)$ for $N=2$ as $\lambda$, the dispersion can be generically written as  $\lambda\sim |\delta \mathbf q|$ or $\lambda\sim |\delta  \mathbf q|^{1/2}$, in the vicinity of the EP, where $\delta \mathbf q = \mathbf q - \mathbf q_\ast$. According to Eq.~\eqref{eq_ofdH}, the dispersion then takes the form $E\sim |\delta \mathbf q|^{1/2}$ or $E\sim |\delta \mathbf q|^{1/4}$. 

\renewcommand{\arraystretch}{1.5}
\begin{table*}[t]
\centering
\begin{tabular}{ |p { 4.7 cm}| p { 6 cm}| p {5.2 cm}|  }
\hline
\multicolumn{3}{|c|}
{Different types of EPs for $N=2$} \\
\hline
\hspace*{ 0.8 cm} $SU(2)$ doublet of EP$_2$ & 
\hspace*{ 2.75 cm} EP$_4$ &\hspace*{ 2.2 cm} EP$_3$ \\ 
\hline
\hspace*{ 1.0 cm}
 $\mathbf B=0$, $\mathbf B'\propto \mathbb I $ 
&  \hspace*{ 1 cm}
$\dim(\ker\mathbf B)+\dim(\ker\mathbf B')=1$,
\newline 
\hspace*{ 1.4 cm}
$\ker(\mathbf B  \, \mathbf B')=
\mathrm{im} ( \mathbf B  \, \mathbf B' )$ &  
 \hspace*{ 0.25 cm}
$\dim(\ker \mathbf B)=\dim(\ker \mathbf B')=1$,
\newline 
 \hspace*{ 0.001 cm} $\textrm{im} \,(\mathbf B')=\ker (\mathbf B)$,
$\textrm{im} \,(\mathbf B) \ne \ker (\mathbf B') $\\
\hline
\hspace*{ 0.2 cm}
 $H(\mathbf q_\ast) =  \mathrm{diag}\{J_2(0), J_2(0)\} $ 
&  \hspace*{ 1.5 cm}
$H(\mathbf q_\ast) =  J_4(0) $
 &  
 \hspace*{0.5 cm}
$H(\mathbf q_\ast) = \mathrm{diag}\{J_3(0), 0\} $\\
\hline
\vspace*{0.01 cm}
\hspace*{ 0.2 cm}
$  \mathbf B=\begin{pmatrix}
0 & 0\\
0 & 0 \\
\end{pmatrix}$,
$ \mathbf B'=\begin{pmatrix}
b' & 0\\
0 & b' \\
\end{pmatrix}$ 
\newline 
& \vspace*{0.01 cm}
\hspace*{ 0.1 cm}
$\mathbf B'
= \begin{cases}
\begin{pmatrix}
b'_1 & b'_2\\
b'_3 &0 \\
\end{pmatrix}, 
& \textrm{when }
\mathbf B=\begin{pmatrix}
0 & 0\\
0 & b_4 \\
\end{pmatrix} \\
& \\
\begin{pmatrix}
b'_1 & b'_2\\
0 & b'_4 \\
\end{pmatrix}, 
& \textrm{when }  \mathbf B=\begin{pmatrix}
0 & b_2\\
0 & 0 \\
\end{pmatrix}
\end{cases} $
&
\vspace*{0.01 cm}
\hspace*{ 0.2 cm}
$    \mathbf B=\begin{pmatrix}
   p_1 \,u_1 & p_1 \,u_2\\
   p_2 \,u_1 & p_2 \,  u_2
\end{pmatrix}$,
\newline 
\newline
\hspace*{ 0.2 cm}
$  \mathbf B'=\begin{pmatrix}
  u_2 \,p'_2 & -u_2 \,p'_1\\
  -u_1 \,p'_2 & u_1\, p'_1
\end{pmatrix}$ 
\newline
\\
\hline
\end{tabular}
\caption{Explanation of the conditions for the existence of different types of EPs when $N=2$. The forms of the matrices $\mathbf B$ and $\mathbf B'$ at the degenerate point $\mathbf q = \mathbf q_\ast$ are shown. Since there is an obvious symmetry under the exchange $\mathbf B \leftrightarrow \mathbf B'$, every case displayed in the table has a $\mathbf B\leftrightarrow \mathbf B'$ partner. The parameters in the bottom row need to further satisfy (1) $b'\ne 0$ in the first column; (2) $\det(\mathbf B')\ne 0$ and $  \mathbf B\ne 0$ in the second column; (3) $|u_1|^2+|u_2|^2\ne 0 $, $|p_1|^2+|p_2|^2\ne 0$, $|p'_1|^2+|p'_2|^2\ne 0$, and $  p'_1 \,p_2-p'_2 \,p_1\ne 0$ in the third column. 
\label{tb_EPSpace}}
\end{table*}

After defining the model, our goal is to work out the Hamiltonian along with the eigenvectors at $E=0$, as well as the nontrivial generalized eigenspace $\mathcal L_0$. At an $n^{\rm{th}}$-order EP, a series of vectors $\{e_0, \,e_1,\,e_2, \,\dots, \,e_n\}$ satisfies the chain equations $H \,e_j=e_{j-1}$, with $e_0$ denoting the null vector and $e_1$ the eigenvector (for $E=0$). When there is no symmetry, the corresponding parameter space of the Hamiltonian, denoted by $\mathcal{EP}_n$, can be figured out easily using the standard methods \cite{PhysRevA.102.032216} (cf. Appendix~\ref{secsupp3}). However, in the presence of sublattice symmetry, employing the standard formalism usually becomes complicated, because it is difficult to find out all the matrices that commute with both the Jordan decomposition and the symmetry transformation.
To avoid this issue, we instead employ the decomposition of each eigenstate as $e_j=(  \psi^T _j, \, \chi^T _j)^T$, such that the condition for the existence of a higher-order EP simplifies to $(i\,\mathbf B\,\chi_j, 
-i\,\mathbf B' \,\psi_j)=(\psi_{j-1}, \,\chi_{j-1})$. As we have already shown that $e_1$ is related to the kernels of $\mathbf B$ and $\mathbf B'$, the chain equations can be solved step by step. The condition for the existence of an EP requires a series of relations between the images $\lbrace \mathrm{im}(\mathbf B),\, \mathrm{im}(\mathbf B') \rbrace $ and the kernels  $\lbrace \ker(\mathbf B), \,\ker(\mathbf B') \rbrace$. From these algebraic relations, we can explicitly work out $\mathcal{EP}_n$. The results are summarized in Table~\ref{tb_EPSpace} (with the derivation shown in Appendix \ref{secsupp1}). We can clearly infer that the results in Table~\ref{tb_EPSpace} cannot be obtained from solutions of some simple continuous equations derived from the Hamiltonian. Hence, a non-Hermitian system exhibits a much richer structure for degeneracies, compared to a Hermitian degeneracy, as observed in the case of no symmetry \cite{PhysRevA.102.032216}. 


\section{The eigenvector structures of different types of EPs for \texorpdfstring{$SU(2)$}{SU(2)}}
\label{sc_eq2d}

In the following subsections, we discuss the properties of various possible EP$_n$'s in great detail, especially focussing on the analytic solutions for the eigenvectors. The system with $N \geq 2$ can host both EP$_2$'s and higher-order EPs, which we discuss below on a case-by-case basis for $N=2$. We denote the location of an EP by $ \mathbf q = \mathbf q_\ast$, and use $\delta \mathbf q=\mathbf q-\mathbf q_\ast$ to parametrize the momentum coordinates in the vicinity of this point. The angle between $\delta \mathbf q$ and the $q_x$-axis is denoted as $\theta$. In other words, near the degenerate point, we parametrize the momentum by $\delta\mathbf q=|\delta \mathbf q| \left ( \cos \theta \, \hat{\mathbf x}+\sin \theta \,\hat{\mathbf y} \right  )$. 
The real parts of the eigenvalues around various kinds of EPs are shown schematically in Fig.~\ref{fig:disp}.
The explicit derivations for the eigenvectors of the higher-order EPs have been worked out in Appendix~\ref{secsupp2}.

\subsection{Lowest-order EPs}

EP$_2$'s are obtained where there is an $SU(2)$ symmetry relating the two flavours of fermions. 
Hence, there must be a $2\times 2$ sub-Hamiltonian that describes a single fermion flavour, and is similar to a 2d Jordan block at the EP.
The full Hamiltonian in Eq.~\eqref{eq_NHYL} at $\mathbf q = \mathbf q_\ast$ is therefore similar to a matrix with two $J_2(0)$ Jordan blocks in the diagonal:
$V \,H(\mathbf q_\ast) \,V^{-1}= 
\textrm{diag}\{
J_2(0),J_2(0) 
\}$.
 On the other hand, the $SU(2)$ symmetry among the two fermion flavours requires the off-diagonal blocks, $\mathbf B(\mathbf q)$ and $\mathbf B'(\mathbf q)$, to be proportional to the identity matrix.
 Hence, at $\mathbf q = \mathbf q_\ast$, $\mathbf B(\mathbf q_\ast) =  0$ (also see the first column of Table~\ref{tb_EPSpace}).  This is a doublet of EP$_2$'s and, to leading powers in $\delta \mathbf q$, the off-diagonal matrices can then be approximated as
\begin{align}
 \mathbf{B}(\mathbf{q}) \simeq
v(\theta)  \, |\delta \mathbf{q}| \,  \mathbb  I_2  \,,\quad 
-i\,\mathbf{B}'(\mathbf{q})\simeq
c  \, \mathbb  I_2 ,\label{eq_doubep2}
\end{align}
where $c$ is a constant. Without any loss of generality, we can parametrize $v(\theta)=v_x \cos \theta +i \,v_y \sin \theta$ \footnote{One can perform a linear coordinate transformation $(\delta q_x,\, \delta q_y)
\to(\delta q'_x,\,  \delta q'_y)$, such that $\mathbf B(\mathbf q) \rightarrow
\mathbf B(\mathbf q') \simeq \mathbb  I_2\otimes v'\, \delta q'$ is holomorphic in the complex coordinate defined as $\delta  q'\equiv \delta q'_x+i\,\delta q'_y$.}, with $v_x$ and $v_y$ being its real and imaginary parts, respectively. The eigenvalues of the resulting Hamiltonian are $\pm \sqrt{c \,v(\theta)\,|\delta \mathbf q|}$, each having a two-fold degeneracy. The four eigenvectors around a doublet of EP$_2$'s are given by $(\pm\sqrt{|\delta \mathbf q|/v(\theta) }, \,0, \, 1, \, 0)^T$ and
$(0,\,\pm\sqrt{|\delta \mathbf q|/v(\theta) },\,0,\,1,)^T$. They coalesce into two linearly-independent vectors as $|\delta \mathbf q|\to 0$. This serves as a typical example of a compound EP, with two EP$_2$'s appearing at $ \delta \mathbf q = 0 $, because each fermion flavour corresponds to a 2d Jordan block at $\mathbf q = \mathbf q_\ast$.


\subsection{Highest-order EPs}

The system supports higher-order EPs once we couple the two different fermion flavours together, and break the $SU(2)$ symmetry. EP$_{4}$'s are the highest-order EPs that can appear, because we have a four-band system.

Because of the sublattice symmetry, the eigenvalues come in pairs of $\lbrace E, -E \rbrace$ --- this implies that the EP$_{4}$ can only appear at $E=0$. Since we require all the eigenvectors to collapse into one at the EP$_{4}$, with $E=0$ being a four-fold degenerate eigenvalue, this brings about several restrictions. First of all, $\lambda=0$ must be a two-fold degenerate eigenvalue of the $2\times 2$ matrix $\mathbf B(\mathbf q_\ast) \cdot \mathbf B'(\mathbf q_\ast)$. Secondly, this matrix product can have only one linearly independent eigenvector. Following the discussion in Sec.~\ref{sc_smEP}, the zero-energy eigenvectors of the Hamiltonian are given by the kernels of $\mathbf B$ and $\mathbf B'$. The single-eigenvector condition thus requires that the total dimension of the kernels, $\dim[\ker \mathbf B(\mathbf q_\ast)]+\dim[\ker \mathbf B'(\mathbf q_\ast)]$, be equal to $1$. Without any loss of generality, we can assume $\dim[\ker \mathbf B(\mathbf q_\ast)]=1$ and $\dim[\ker \mathbf B'(\mathbf q_\ast)]=0$. If we denote the zero-energy eigenstate of $\mathbf B(\mathbf q_\ast)$ as $\chi_1$, the four-dimensional generalized eigenspace $\mathcal L_0$ of $H(\mathbf q_\ast)$ has the first vector $e_1$ proportional to $(0, \,\chi_1^T )^T$. The details of sorting out this generalized eigenspace have been explained in Appendix~\ref{secsupp1}.

\begin{figure}
    \centering
 \subfigure[]{\includegraphics[width=0.38\linewidth]{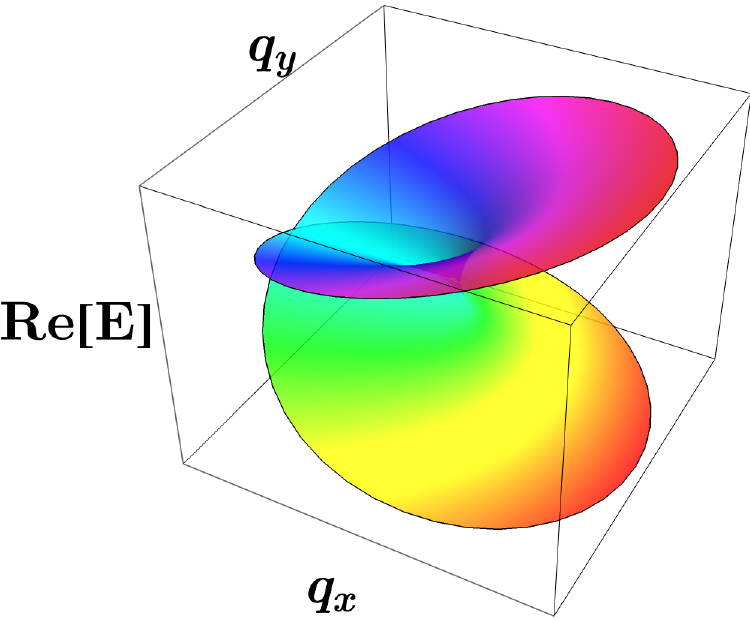}}\quad
  \subfigure[]{\includegraphics[width=0.38\linewidth]{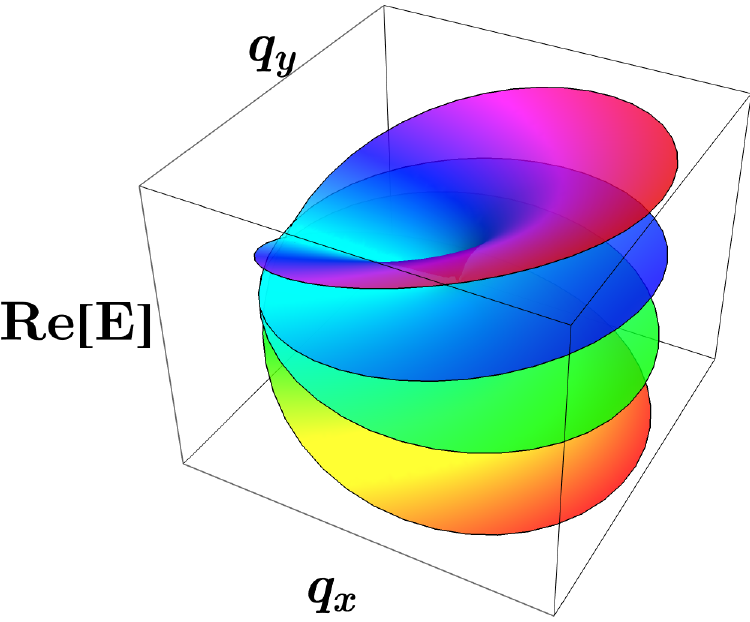}}\quad
 \subfigure[]{\includegraphics[width=0.38\linewidth]{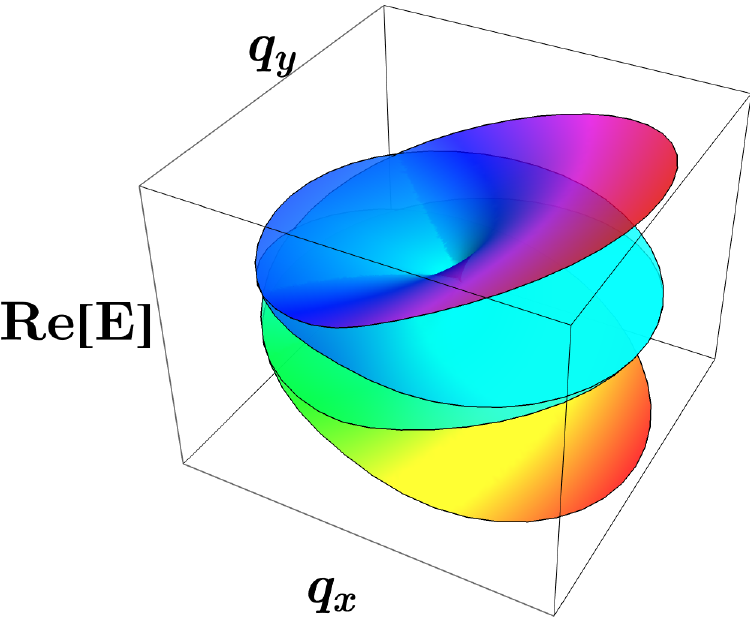}}
    \caption{Real parts of the eigenvalues $E$ for the cases of different types EPs when $N=2$: (a) $\Re[E]$ for a doublet of EP$_2$'s, where each eigenvalue is doubly degenerate. (b) $\Re[E]$ for an EP$_4$, where all different energy eigenvalues coalesce at one singular point. (c) $\Re[E]$  for an EP$_3$, for which the eigenvectors are sensitive to how the point of singularity is approached in the Brillouin zone. We choose the EP to be anisotropic. The scaling of $E$ around the EP can take different forms along different directions.}
    \label{fig:disp}
\end{figure}

The EP$_4$ Hamiltonian at $\mathbf q_\ast$ is similar to a four-dimensional Jordan block, i.e.,
$V \,H(\mathbf q_\ast) \,V^{-1}= J_4 (0)$.
We present a concrete example, which follows the forms shown in the second column of Table~\ref{tb_EPSpace}, by turning on the minimal number of non-Hermitian hoppings. To leading power in $|\delta \mathbf q|$,
\begin{align}
    \mathbf B(\mathbf q_\ast+\delta \mathbf q)&\simeq\begin{pmatrix}
 v_1(\theta)\, |\delta \mathbf q|  & b_2 \\
0 & v_4(\theta) \,|\delta \mathbf q|  \\
\end{pmatrix},
\nonumber \\  
\mathbf B'(\mathbf q_\ast+\delta \mathbf q)&\simeq\begin{pmatrix}
 b'_1  & 0 \\
  v'_3(\theta)\, |\delta \mathbf q| & b'_4 \\
\end{pmatrix} ,
\end{align}
where $b_j$ and $b'_j$ are constants, and $v_j(\theta)$ and $v'_j(\theta)$ are functions of the angle $\textrm{arg}(\delta q_x + i\,\delta q_y)$. More precisely, we assume that these parameters contain ${\mathcal O} (|\delta \mathbf q|)$ corrections, so that we do not lose crucial terms when expanding our eigenvalues and eigenvectors in powers of $|\delta \mathbf q|$. Using Eq.~\eqref{eq_ofdH}, the eigenvalues and the eigenstates are given by (more details can be found in Appendix~\ref{secsupp2})
\begin{align}
 &E= {\mathcal O} (\sqrt{|\delta \mathbf q|})\nonumber\\ \text{ and }&
 e= \left  ({\mathcal O} (|\delta \mathbf q|^{1/2}),
 \, {\mathcal O} (|\delta \mathbf q|^{3/2}), \, 1, \, {\mathcal O} (|\delta \mathbf q|) \right )^T,
\end{align}
respectively.
The eigenvalues vanish as $\sqrt{|\delta \mathbf q|}$ [cf. Fig.~\ref{fig:disp}(b)], while the four eigenvectors converge to $\left (0, \,0, \,1,  \,0 \right)^T$, right at the EP. Although the dispersions scale as square roots (rather than quartic roots) around the EP$_4$, the typical behaviour of an EP$_4$ involving the eigenvector-coalescence into a single one is observed.

We would like to emphasize that the EP$_4$ here does not exhibit a quartic-root dispersion. This is expected as an EP$_n$ can exhibit arbitrary $m^{\text{th}}$-order root singularity, where $m\le n$ \cite{demange2011signatures,sayyad2022realizing}, or even dispersions that cannot be expressed as root functions \cite{konig2022braid}. In Appendix~\ref{sec_suppEP$_4$}, we show an example where a singularity in the form of a root of quartic order is realized in our four-band sublattice-symmetric system.

\subsection{Odd-order EPs}
\label{sc_oddEP}

As we have shown in Sec.~\ref{sc_smEP}, the sublattice symmetry requires the dispersion near an EP at $E=0$ to scale as $\delta E \sim |\delta \mathbf q|^{1/(2p)}$, with $p \in {\mathbb Z}^+$. In addition, the sublattice symmetry also restricts the ways in which $E\ne 0$ eigenvectors coalesce. These conditions seem to obstruct an odd-order EP. However, through an explicit construction of an EP$_3$ for the $N=2$ four-band model, we will show that a somewhat anomalous EP$_3$ can exist. Although the generic case is expected to exhibit a cube-root dispersion around the singularity, a sublattice symmetry forces it to have a square-root-dispersion \cite{ips-prl}, which is indeed found to be the case here. We also find that the way the eigenvectors coalesce with one another depends on the path chosen to approach the EP$_3$ (while a regular EP$_3$ has three eigenvectors collapsing together for any path). The EP$_3$ here is anomalous and different from the usual scenarios.

Because of the sublattice symmetry, a zero eigenvalue can appear only with an even algebraic multiplicity. Hence, for the $N=2$ case, the existence of an EP$_3$ with $ E=0$ requires that its algebraic multiplicity must be four. The degenerate point is thus an EP$_3$ plus an accidental zero-energy eigenstate. According to our symmetry analysis, the total dimension of the kernels for $\mathbf B(\mathbf q_\ast)$ and $\mathbf B (\mathbf q_\ast)$ is $m+n=2$. If $m=2$ and $n=0$, the matrix $\mathbf B(\mathbf q_\ast)$ is identically zero, and $\mathbf B'(\mathbf q_\ast)$ can be brought to a diagonal matrix via a transformation matrix $\mathbf V$. Applying the transformation matrix $\textrm{diag}(\mathbf V, \,\mathbf V)$ to $H(\mathbf q_\ast)$ then brings it explicitly to a form similar to Eq.~\eqref{eq_doubep2}. Hence, either $(m=2,\,n=0)$ or $(m=0, \,n=2)$ gives a doublet of EP$_2$'s. An EP$_3$ can emerge only when $m=n=1$.

Now we look at a specific example. According to Table~\ref{tb_EPSpace}, an EP$_3$ appears when
$ V \,H(\mathbf q_\ast) \,V^{-1}= 
\textrm{diag}\{
J_3(0), 0 \} $, and
\begin{align}
    \mathbf B(\mathbf q_\ast)=\begin{pmatrix}
0  & b_2 \\
0   & 0 \\
\end{pmatrix},   
\quad \mathbf B'(\mathbf q_\ast)=\begin{pmatrix}
b'_1  & b'_2 \\
0  & 0  \\
\end{pmatrix}.
\end{align}
There are two linearly independent eigenvectors at $E=0$, which are proportional to $e_1=\left (0,\,0, \,1,\,0 \right )^T$ and $e_2= \left (b'_2, \,-b'_1,\,0,\,0 \right )^T$, proving that it is {\it not} an EP$_4$. From the Jordan decomposition, we find that $e_1$ belongs to a generalized eigenspace of dimension three, such that $e_1=H(\mathbf q_\ast) \,\tilde e_2$ and $\tilde e_2=H(\mathbf q_\ast) \,\tilde e_3$, with $\tilde e_2= \left (1/b'_1,\,0,\,0,\,0 \right )^T$ and $\tilde e_3=\left (0,\,0, \,0,\,1/(b_2 \,b'_1) \right )^T$. Hence, this is an EP$_3$ accidentally coinciding with a zero-energy eigenvector.

To investigate how the symmetry constraints play out in this case, we explicitly show how the eigenvectors behave in the vicinity of this EP$_3$. As the sublattice symmetry forbids the three eigenvectors folding together, they show an anomalous behaviour, which is in-between the coaslescence features of the eigenvectors of EP$_2$ and EP$_4$. This is the reason why the eigenvector-coalescence depends on the path chosen while approaching $ \mathbf q_\ast$. Whenever an EP is anisotropic \cite{PhysRevB.99.241403}, the eigenvectors indeed exhibit an enhanced path-dependent sensitivity.

\begin{figure*}
    \centering
 \subfigure[]{\includegraphics[width=0.301 \linewidth]{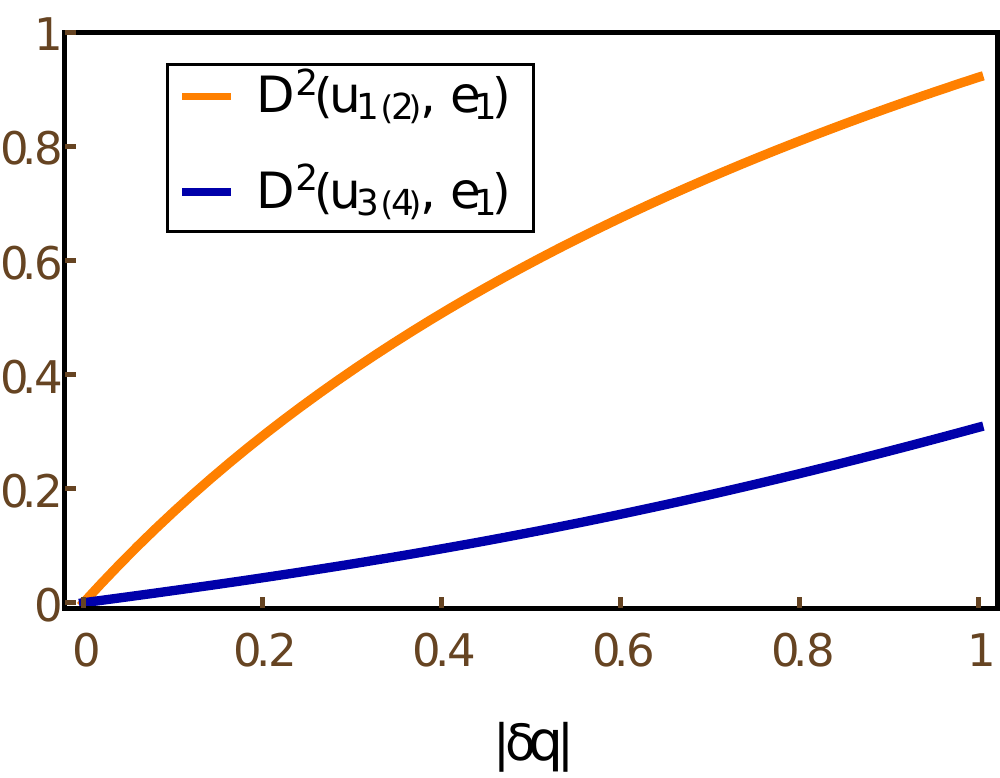}}\qquad
  \subfigure[]{\includegraphics[width=0.3\linewidth]{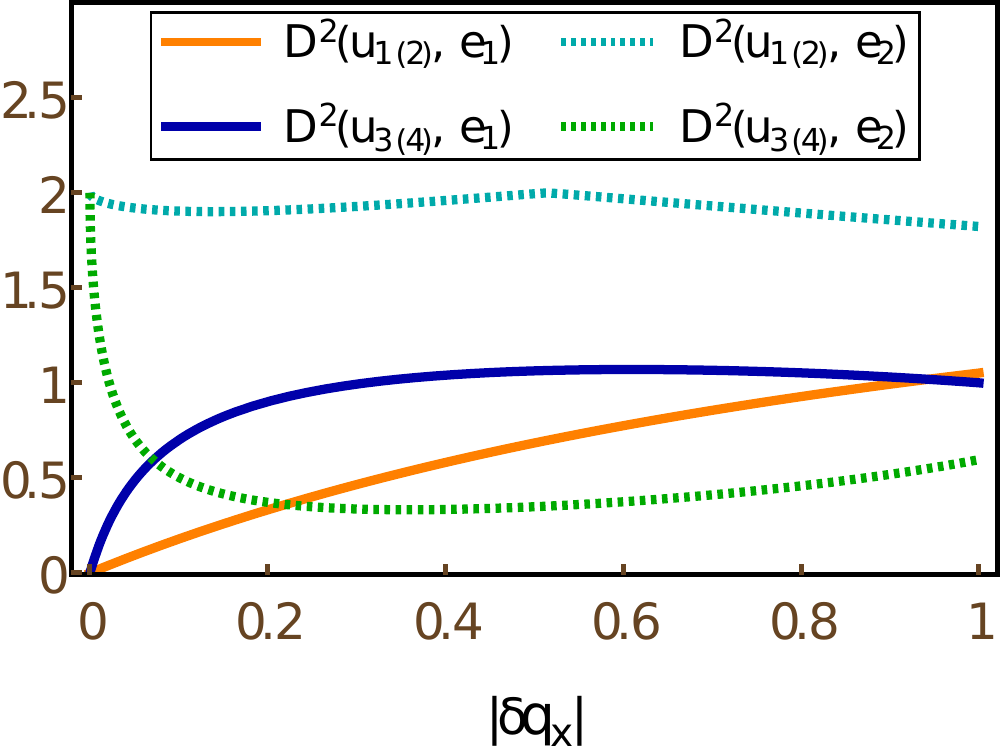}}\qquad
 \subfigure[]{\includegraphics[width=0.3\linewidth]{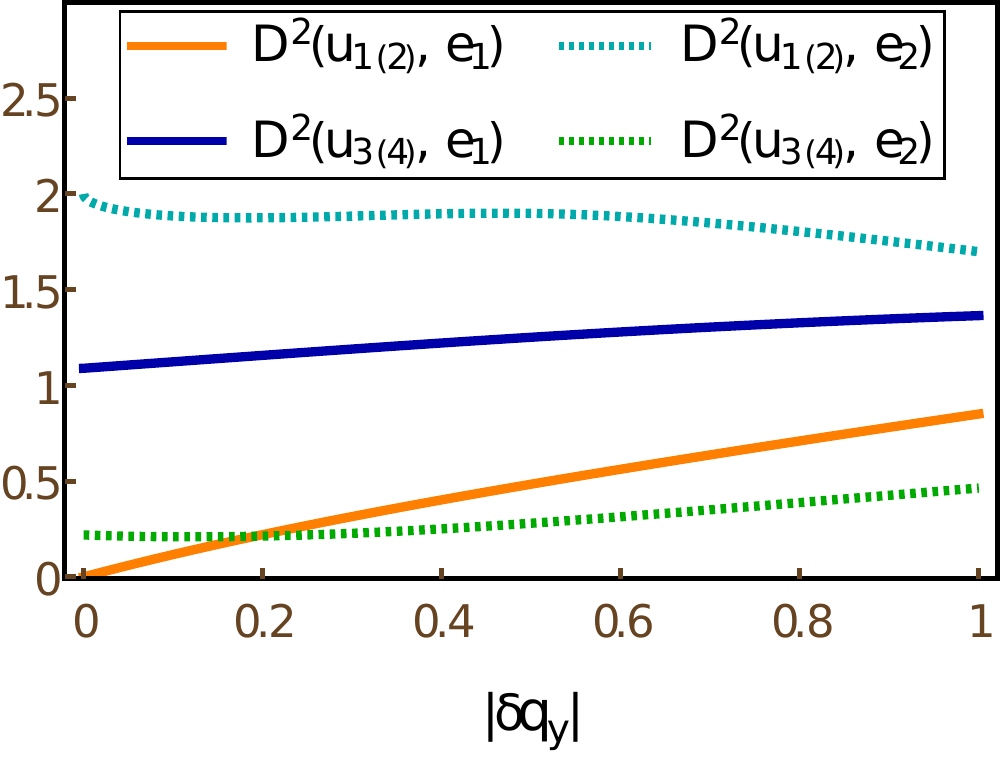}}
\caption{\label{fig_EPd}
Square of the quantum distance ($D^2$) describing how eigenstates coalesce into those at the exceptional points of different orders: (a) $D^2(u_i, e_1)$ goes to zero at the EP$_4$ for all $i \in[1,4]$.
(b) Approaching the EP$_3$ along the {\it Path 1} of Sec.~\ref{sc_oddEP} (with $q_y=0$), for all $i \in[1,4]$
$D^2(u_i, e_1)$ goes to zero, while $D^2(u_i, e_2)$ goes to some nonzero values.
(c) When one approaches the EP$_3$ along the {\it Path 2} of Sec.~\ref{sc_oddEP} (with $q_x =0$), $D^2(u_1,e_1)$ and $D^2(u_2,e_1)$ go to zero, while $D^2(u_3,e_j)$ and $D^2(u_3,e_j)$ remain nonzero for both $j=1, 2$.}
\end{figure*}

\subsubsection{Path 1}

 We approach the EP along the $q_x$-direction (i.e., $ q_y =0$ along this path), assuming that all deviations are linear, in which case the expansion looks like
\begin{align}
   \mathbf B(\mathbf q_\ast+ \delta  q_x \, \hat{\boldsymbol x})
 &\simeq
      \begin{pmatrix}
 v_1(0) \,|\delta  q_x|  & b_2 \\
v_3(0)\, |\delta  q_x|  & v_4(0) \, |\delta  q_x|  \\
\end{pmatrix},\nonumber \\
\mathbf B'(\mathbf q_\ast+ \delta  q_x \, \hat{\boldsymbol x})&\simeq
\begin{pmatrix}
 b'_1  & b'_2 \\
v'_3(0) \, |\delta  q_x|  & v'_4(0) \,|\delta  q_x|  \\
\end{pmatrix}.
\label{ep3case1}
\end{align}
As before, we implicitly assume that the variables $\lbrace b_j,\,b'_j \rbrace $ and $\lbrace v_j(0), \,v'_j(0) \rbrace $ can contain ${\mathcal O} (|\delta  q_x|)$ corrections. The product
\begin{align}
    &\mathbf B'(\mathbf q_\ast+ \delta  q_x \, \hat{\boldsymbol x})\cdot
    \mathbf B(\mathbf q_\ast+ \delta  q_x \, \hat{\boldsymbol x})\nonumber\\
    \simeq &
  \begin{pmatrix}
 [b'_1 \,v_1(0)+b'_2 \,v_3(0)] \,|\delta  q_x|  & b_2 \,b'_1 \\
[v_1(0) \,v'_3(0)+v_3(0) \,v'_4(0)] \,|\delta  q_x|^2  & b_2 \,v'_3(0) \,|\delta  q_x|  \\
\end{pmatrix}
\end{align}
determines the eigenvalue $E^2$ and $\chi$ [cf. Eq.~\eqref{eq_ofdH}]. 
The two eigenvalues of the above matrix vanish as ${\mathcal O} (|\delta q_x|)$, while its two eigenvectors approach $(1,0)^T$ as $\left (1,\,{\mathcal O} (|\delta q_x|) \right )^T$ (the intermediate steps are shown in Appendix~\ref{secsupp2}). Hence, the deviation in dispersion scales as $\delta E\sim \sqrt{|\delta q_x|}$. Since the upper component is given by $\psi= i \, \mathbf B \, \chi/E$, it vanishes as $\left ({\mathcal O} (\sqrt{|\delta q_x|}), \,{\mathcal O} (\sqrt{|\delta q_x|})  \right )^T$. Therefore, all the four eigenvectors coalesce to $e_1=(0,\,0, \,1,\,0)^T$ at $\mathbf q = \mathbf q_\ast$. In comparison, there is no eigenvector converging to the eigenvector $e_2$ at $\mathbf q_\ast$. Although this EP is of order three, its singularity behaviour along the $q_x$-path is similar to a typical EP$_4$.

\subsubsection{Path 2}

The eigenvectors exhibit a typical EP$_2$ behaviour if $v'_3(\theta)$ and $v'_4(\theta)$ vanish for some angle 
$\theta$, which can be obtained by imposing an additional symmetry to these parameters. For convenience, we choose the direction of approach to the EP in this case to be along the $q_y$-direction, and set $v'_3(\pi/2)= v'_4(\pi/2)=0$. The off-diagonal matrices take the forms:
\begin{align}
      \mathbf B(\mathbf q_\ast+  \delta  q_y \, \hat{\boldsymbol y})&\simeq
      \begin{pmatrix}
 v_1(\pi/2) \,|\delta q_y|  & b_2 \\
v_3(\pi/2) \,|\delta q_y|  & v_4(\pi/2)\, |\delta q_y|  \\
\end{pmatrix},
\nonumber \\
\mathbf B'(\mathbf q_\ast+ \delta  q_y \, \hat{\boldsymbol y})&\simeq
\begin{pmatrix}
 b'_1  & b'_2 \\
r'_3 \, |\delta q_y|^2  & r'_4 \,|\delta q_y|^2  \\
\end{pmatrix},
\label{ep3case2}
\end{align}
and their product is given by
\begin{align}
     &\mathbf B'(\mathbf q_\ast+ \delta  q_y \, \hat{\boldsymbol y})\cdot\mathbf 
    B(\mathbf q_\ast+ \delta  q_y \, \hat{\boldsymbol y})\nonumber\\
    \simeq &
  \begin{pmatrix}
 \left [b'_1 \,v_1(\pi/2)+b'_2 \,v_3(\pi/2)  \right ] |\delta q_y|  & b_2 \,b'_1 \\
\left [ v_1(\pi/2)\, r'_3+v_3(\pi/2)\,r'_4  \right ] |\delta q_y|^3  
& b_2 \,r'_3\,|\delta q_y|^2  \\
\end{pmatrix}.
\end{align}
One of its eigenvalues of the product matrix vanishes as $\lambda_1={\mathcal O} (|\delta q_y|)$, while the other vanishes as $\lambda_2={\mathcal O} (|\delta q_y|^2)$ (the derivations are shown in Appendix~\ref{secsupp2}). Note that this gives a different scaling for the dispersion around the EP, compared to the {\it Path 1}. The two eigenvectors of $\mathbf B'\cdot\mathbf B$ behave as $\chi_1 \simeq(1,\, {\mathcal O} (|\delta q_y|^2))^T$ and $\chi_2\simeq \left (1,   {\mathcal O} (|\delta q_y|) \right )^T$, respectively. The corresponding upper components (obtained from the relations $\psi_a = i \,\mathbf B \,\chi_a /E$, with $a \in \lbrace 1, 2\rbrace $) thus scale as $\psi_1 \sim \left ({\mathcal O} (\sqrt{|\delta q_y|}),  {\mathcal O} (\sqrt{|\delta q_y|}) \right )^T $ and $\psi_2\sim  \left  ( {\mathcal O} (1),  {\mathcal O}(1) \right )^T $, respectively. In this situation, the two eigenvectors $(\pm\psi^T_1, \chi^T_1)$ converge to $e_1^T$, while the other two eigenvectors $(\pm\psi^T_2,  \chi^T_2)$ go to two other linearly-independent vectors, which we denote as $e_3^T$ and $e_4^T$. Hence, along this path, the eigenvectors behave as a single eigenvector of an EP$_2$ plus two linearly-independent accidental zero-energy eigenvectors.

\section{Irregular subspace topology of the EPs}
\label{sc_todis}

In this section, we will formulate a way to quantitatively characterize the overlap of eigenvectors, following which we will illustrate the origin of the anomalous behaviour of the odd-order EPs under sublattice symmetry. The conclusion that comes out of this set-up is that eigenvector-coalescence is not actually a point-like property of the EP itself, but it depends on how the Hamiltonian looks like in its neighborhood. In fact, we will see that for our example of $N=2$, the EP$_3$ under sublattice symmetry can in fact be understood as the point at which the parameter spaces of EP$_4$ and EP$_2$ intersect. This feature comes from the subspace topology of $\mathcal{EP}_n$, as a subspace of all $4\times 4$ matrices $M_4(\mathbb C)$.

When analyzing the coalescence of eigenvectors, it can be ambiguous if we directly compare them, because
eigenvectors are equivalent upto phases. In order to characterize unambiguously how the states coalesce near regular EPs and mixed-type EPs, it is most convenient to introduce the quantum distance $D$ \cite{provost1980riemannian}, such that
\begin{align}
    D^2(u,u') & =\underset{ \lbrace \alpha, \beta \rbrace \in \,\mathbb R}  
    {\textrm{inf }}  \, \vert\vert \, u\, e^{i\,\alpha}
 - u' \,e^{i\,\beta}\vert\vert^2 \nonumber \\
 & = 2-2\, \vert\langle u\,| \,u'\rangle\vert\,.
\end{align}
Clearly, $D^2(u, u')$ is invariant under $U(1)\times U(1)$ transformations, i.e., under the change of the phases of $u$ and $u'$. 
Here the states are normalized as $\langle u \,| \,u\rangle=\langle u'\,| \,u'\rangle=1$, and $||\cdot|| $ is the usual norm $\sqrt{ \langle \cdot|\cdot\rangle }$ of a quantum state. Using $u'$ to denote the eigenvectors at the EP at $ \mathbf q  = \mathbf q_\ast$, and $ u $ to denote the states away from $\mathbf q_\ast $, $D$ is a function of $( \mathbf q-\mathbf q_\ast )$. $D^2$ is positive-definite, and vanishes only when $u$ and $u'$ differ by a phase (i.e., when $u$ and $u'$ denote the same quantum state). Hence, $D^2(u, e_j)$ can be used to describe unambiguously how the eigenvectors are approaching their target eigenvectors at the EP. 

Since the eigenstates $e_j$'s at the EP (i.e., at $E=0$) are invariant under the sublattice symmetry, the two nondegenerate eigenstates $(\pm\psi,\chi)$, related by the sublattice symmetry, have the same $D^2$ value with $e_j$. In Fig.~\ref{fig_EPd}, we show how the eigenvectors approach the ones at the EPs, as $\mathbf q$ approaches $\mathbf q_\ast$. In all the cases, the four nondegenerate states fall into two classes: each corresponding to a sublattice symmetry-related pair. Let us denote the two pairs of eigenvectors as $\lbrace u_1, u_2 \rbrace $ and $\lbrace u_3, u_4 \rbrace $.
For the EP$_4$, $D^2$ is computed from $e_1$ (which is the sole linearly-independent eigenvector right at the EP) and each of the four nondegenerate eigenvectors, and it goes to zero as we approach the EP$_4$. However, things are more complicated for the EP$_3$, and in fact the behaviour of $D^2$ corroborates the results obtained in Sec.~\ref{sc_oddEP}.
Approaching the EP along the \textit{Path 1} of Sec.~\ref{sc_oddEP} (with $q_y=0$), for all $i \in[1,4]$, $D^2(u_i, e_1)$ goes to zero, while $D^2(u_i, e_2)$ remains nonvanishing.
On the other hand, if one approaches the EP along the \textit{Path 2} (with $q_x =0$), $D^2(u_1,e_1)$ and $D^2(u_2,e_1)$ go to zero, while $D^2(u_3,e_j)$ and $D^2(u_3,e_j)$ remain nonzero for both $j=1$ and $j=2$.

\begin{figure*}
    \centering
\includegraphics[width=0.7 \linewidth]{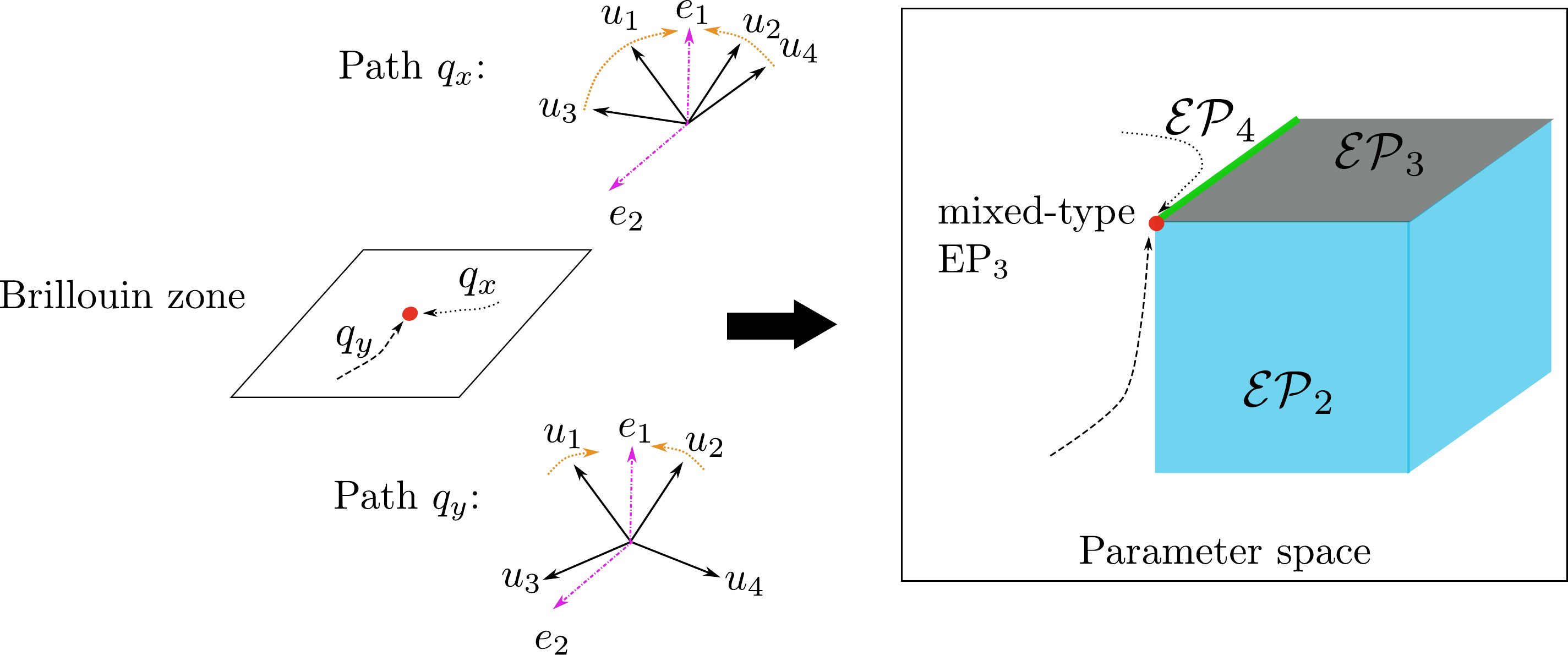}
  \caption{Schematic depiction of the location of a mixed-type EP$_3$ in the parameter space of a non-Hermitian matrix. The white (uncoloured) region in the parameter space represents matrices with nondegenerate eigenvalues. They are dense and their parameter space has the highest dimensionality. In the absence of any symmetry, the dimension of the  $\mathcal{EP}_n$ space decreases as $n$ increases. The mixed-type EP$_3$ appears as the intersection point of the $\mathcal{EP}_2$ (light blue cube), $\mathcal{EP}_3$ (gray surface), and $\mathcal{EP}_4$ (green line). When the sublattice symmetry is imposed, the regular
$\mathcal{EP}_3$ surface (gray region) is forbidden, and the mixed-type EP$_3$ can be approached only via the neighbourhood of either $\mathcal{EP}_4$ (dotted line) or $\mathcal{EP}_2$ (dashed line). This leads to two different ways of eigenvector-coalescence, which are shown by the collapse of directed arrows against ``Path $q_x$'' and ``Path $q_y$'' (corresponding to {\it Path 1} and {\it Path 2} of Sec.~\ref{sc_oddEP}, respectively).
\label{fig_EPsp}}
\end{figure*}

The anomalous behaviour of the eigenvectors near the EP$_3$ can be explained by its mixed nature. This is a very special property of a Jordan decomposition when the diagonal of a Jordan block coincides with some other eigenvalue(s). An EP with such a Jordan block is qualitatively different from an EP whose Jordan block has a nonzero gap with other eigenvalues. We denote the latter as regular EPs. The space $\mathcal{EP}_3$ comprises two sets, namely, the set $U_1$ of regular EP$_3$ and the set $U_2$ of mixed-nature EP$_3$.

To illustrate the possible structures around an EP$_3$, we consider a $4\times 4$ matrix $M$ with no particular symmetry. Such a matrix has a $16$ (complex) dimensional parameter space $M_4(\mathbb C)$. The most common matrices in this space are those which are the non-singular ones featuring nondegenerate eigenvalues. The EPs are represented by matrices with singularity, and they form lower-dimensional subspaces of $M_4(\mathbb C)$. The dimension of the parameter space $\mathcal{EP}_n$ decreases as $n$ becomes larger (see Appendix~\ref{secsupp3}). For an EP$_3$ with Jordan decomposition
$ M \big \vert_{\rm{EP}_3}
=\mathrm{diag}\lbrace J_3(0), \,0 \rbrace$,
one can easily verify that, within $\mathcal{EP}_2$,
there is a sequence of points whose limit is $ M\big \vert_{\rm{EP}_3}$:
\begin{align}
\lim \limits_{\epsilon\to 0}&
\left[
 M(\epsilon) \big  \vert_{\rm{EP}_2} =
\begin{pmatrix} 0 & 1 & 0 & 0  \\ 0 & 0 & 1 & 0 \\ 0 & 0 & \epsilon & 0 \\ 0 & 0 & 0 & 2\epsilon \\ \end{pmatrix}\in \mathcal{EP}_2\right]
=  M \big \vert_{\rm{EP}_3}\nonumber\\
=&\begin{pmatrix}
 0 & 1 & 0 & 0  
 \\  0 & 0 & 1 & 0 \\ 0 & 0 & 0 & 0 \\ 0 & 0 & 0 & 0 \\ 
 \end{pmatrix}
 \in U_2\subset\mathcal{EP}_3 \,. 
 \end{align}
This implies that, in any neighbourhood of $  M \big \vert_{\rm{EP}_3} $, we can always find points belonging to $\mathcal{EP}_2$. In particular, when the matrices representing nondegenerate eigenvalues are close enough to the matrices representing EP$_2$'s, two of the four eigenvectors of our non-Hermitian Hamiltonian should also come close to each other (see Fig.~\ref{fig_EPsp}).

In addition to the above limit, we can find another limit by tuning the parameters of the matrix containing the Jordan block of the EP$_3$, such that the EP$_3$ of $M\big \vert_{\rm{EP}_3}$ is now a limiting point of an $\mathcal{EP}_4$. This can be seen from
\begin{align}
\lim \limits_{\epsilon\to 0}&\left[  M(\epsilon) \big  \vert_{\rm{EP}_4 }
= \begin{pmatrix} 
0 & 1 & 0 & 0  \\ 0 & 0 & 1 & 0 \\ 0 & 0 & 0 & \epsilon \\ 0 & 0 & 0 & 0 \\ 
\end{pmatrix}\in 
\mathcal{EP}_4\right]
= M \big \vert_{\rm{EP}_3}\nonumber\\
=&\begin{pmatrix} 
0 & 1 & 0 & 0  \\  
0 & 0 & 1 & 0 \\ 
0 & 0 & 0 & 0 \\ 
0 & 0 & 0 & 0 \\ 
\end{pmatrix}
\in U_2\subset\mathcal{EP}_3 \,. 
\end{align}
This result is much more counter-intuitive than the coincidence with the EP$_2$ case, because $\mathcal{EP}_4$ has a lower dimension than $\mathcal{EP}_3$, and the region $  M(\epsilon) \big  \vert_{\rm{EP}_4 }$ in the neighbourhood of $  M \big  \vert_{\rm{EP}_3 }$ is usually very small. However, the neighbourhood of $\mathcal{EP}_4$ comprising all matrices representing nondegenerate eigenvalues is not small. These matrices then can have a large overlap with the nondegenerate neighbourhood of $  M \big  \vert_{\rm{EP}_3 }$. As a result, all paths through this intersecting region will show a behaviour characteristic of a four-fold eigenvector-coalescence.

The arguments above show that a mixed-type EP can appear as a common limit point of lower- and higher-order EPs, which implies that such an exceptional degeneracy cannot form a closed subspace in $M_4(\mathbb C)$ by simply combining certain higher-order EPs. This anomalous behaviour of the odd-order EPs is absent in Hermitian systems. In the parameter space of a Hermitian matrix $H_{\text{herm}} $, if we denote the space with an $n$-fold degenerate eigenvalue $E$ as $\mathcal{HD}_n(E)$, the space $\cup_{n\ge m} \mathcal{H D}_n(E) $ is given by the zeros of the resultants ($ R$) or discriminants ($\mathcal D$), i.e., by $R(E)=0$ \cite{PhysRevLett.127.186602} or $\mathcal D[ H_{\text{herm}} (E)]=0$ \cite{sayyad2022realizing}. These equations involve continuous functions in $M_4(\mathbb C)$, and hence their solutions constitute a closed subspace of $M_4(\mathbb C)$. This means that the limit of a series Hermitian degeneracy  $\mathcal{HD}_n$ can only end in some $\mathcal{HD}_m$ ($m\ge n$). As a result, only higher-order degeneracy can be the limit of a lower-order degeneracy, but not the other way around. Therefore, for Hermitian matrices, there is no mixed-type degeneracy.

In summary, the enhanced eigenvector sensitivity can be understood intuitively in the following way (see also Fig.~\ref{fig_EPsp}). The different directions of approaching $\mathbf q_\ast$ in the Brillouin zone can be mapped to approaching the EP$_3$ through different tracks in the space of matrices representing nondegenerate eigenvalues. Due to the sublattice symmetry, it is forbidden to approach the EP$_3$ through the neighbourhood of the matrices representing a regular EP$_3$. Consequently, the sublattice symmetry-restricted EP$_3$ can only be reached through the neighbourhoods of $\mathcal{EP}_2$ and $\mathcal{EP}_4$. Of course in those neighbourhoods, either two or four eigenvectors coalesce together, leading to the anomalous behaviour of the eigenvectors of the EP$_3$.

\section{Lattice realizations and expectations for generic $N$-values}
\label{secgen}

Examples of fermionic Hamiltonians with sublattice symmetry include solvable spin liquid models, such as the Kitaev spin liquid \cite{kitaev,kang-emil,yang2022exceptional} (corresponding to $N=1$), and the Yao-Lee $SU(2)$ spin liquid \cite{yao-lee} (corresponding to $N=3$).
The $N=2$ model studied in this paper can be embedded in the Yao-Lee model. There, the low-energy physics is described by $N=3$ flavours of Majorana fermion operators, with the Hamiltonian consisting of only nearest-neighbour hoppings amongst fermions of the same flavour. In order to produce the higher-order degeneracies discussed in this paper, we need to introduce terms which couple different flavours [thus breaking the $SU(N)$ symmetry] -- these can be generated by terms $\sigma_{\alpha,i}\,\tau^x_{i}\,\tau^x_{j}\,\sigma_{\beta,j}$ (with $i\neq j$) in terms of the original spin operators. The details have been outlined in Appendix~\ref{secsupp4}.

Setting $N=3$, one can get EPs up to sixth order, which is then expected to display a richer eigenvector sensitivity. For this case, an EP$_5$ can exist where a five-dimensional Jordan block becomes degenerate with another band. Near this EP$_5$, the coalescence of eigenvectors can be four-fold or six-fold. Moreover, since two-fold coalescence is also permitted by the sublattice symmetry, there exists paths along which the eigenvectors collapse like they do in the vicinity of an EP$_2$. Consequently, such an EP$_5$ has a higher degree of eigenvector sensitivity, making it possible to have more knobs to tune quantum states.

For a generic value of $N$, in order to obtain an $N$-fold compound EP$_2$, or a highest-order simple EP$_{2N}$, the algebraic conditions are simply obtained by replacing the expressions for $N=2$ by the appropriate $N$-value. More specifically, the $N$-fold EP$_2$ is $SU(N)$-invariant, and is obtained by choosing $\mathbf B$ as a diagonal matrix vanishing at $\mathbf q_\ast$, while $\mathbf B'(\mathbf q_\ast)$ remains nonzero. As for EP$_{2N}$, we need $\dim(\ker\mathbf B)+\dim(\ker\mathbf B')=1$ for generic $N$ as well. Additionally, in order to ensure that all the $2N$ linearly-independent eigenvectors coalesce to a single one, we need to impose the condition $\mathbf B\cdot\mathbf B'\sim J_{2N}(0)$, which can alternatively be represented as $\ker(\mathbf B\cdot\mathbf B')^m=\mathrm{im}(\mathbf B\cdot\mathbf B')^{2N-m}$ (with $0<m<2N$). For EPs with orders between $2$ and $2N$, the analysis becomes more complicated. Mixed-type odd-order EPs will exist at $E=0$, analogous to the EP$_3$ of the $N=2$ case that we have explicitly studied. Although the dimensions of the kernels can be worked out in a way similar to that shown in Table~\ref{tb_EPSpace}, the image and kernel relations need to be figured out on a case-by-case basis, and closed-form expressions for the eigenvectors might involve extremely complicated calculations. Nevertheless, the generic topological relations between higher-order EPs remain valid.

\section{Summary and outlook}
\label{secsum}

In this paper, we have explored the emergence of higher-order EPs in two-dimensional four-band non-Hermitian systems, with a sublattice symmetry. Such systems are relevant to non-Hermitian extensions of solvable spin liquid models. The sublattice symmetry forces the eigenvalues to appear in pairs of $\lbrace E,  -E \rbrace $, and the dispersion around an EP is restricted to be an even root of the deviation in the momentum space. We have explicitly computed how the eigenvectors collapse at an EP, and found an anomalous behaviour for odd-order EPs. Based on the analytical solvability of a four-band system, we have shown that the collapse of the eigenvectors depends on the specific path of approaching an EP$_3$. The behaviour is anomalous in the sense that it is in contradiction with the intuition that $n$ eigenvectors always coalesce together at an EP$_n$. In fact, the number of collapsing eigenvectors for a mixed-type odd-order EP is an even number smaller or greater than $n$, which is caused by the presence of the sublattice symmetry. Intuitively, this unconventional feature can be understood from the fact that there is a restriction in the parameter space of EP$_3$ due to the sublattice symmetry, and this unusual EP$_3$ can be approached only via the neighbourhoods of EP$_2$'s and EP$_4$'s.

Using the notion of a quantum distance, we have further explored the behaviour of the eigenvectors near the mixed-type EP$_3$. We have found that the eigenvectors do not necessarily converge to those of a regular EP$_3$, especially when we are approaching it from a neighbourhood of $\mathcal{EP}_2$. The quantum distance to the eigenvectors at the mixed-type EP$_3$ can change abruptly if we slightly perturb the approaching process. It is already known that the non-unitary evolution under a non-Hermitian Hamiltonian leads to a shorter quantum distance \cite{PhysRevLett.98.040403,PhysRevLett.99.130502}, which can play a role in state preparation. Hence, we expect that the anomalous behaviour near higher-order EPs will significantly enhance this effect, and lead to novel applications exploiting the features we have discovered through our analysis.

The enhanced eigenvector sensitivity for the mixed-type EP$_3$s is a reminiscence of generic counter-intuitive features specific to non-Hermitian systems (i.e., these are absent in the corresponding Hermitian counterparts). A very well-known example is the non-Hermitian skin effect \cite{skin-effect,PhysRevLett.124.250402,PhysRevB.99.081302,PhysRevLett.125.118001,weidemann2020topological,xiao2020non,PhysRevLett.123.016805,PhysRevB.102.205118,PhysRevB.106.115107}, where a very small change in the boundary conditions brings about remarkable modifications to the spectrum. The mixed nature of the odd-order EPs also generalizes the notion of the recently-studied non-defective EPs \cite{sayyad2022symmetry}, where a Hermitian degeneracy mixes with the usual EP$_2$.

A promising future research direction is to explore analogous unconventional EPs in three-dimensional systems with appropriate symmetry constraints. The extended dimensionality is expected to provide a richer parameter space for the characterization of generic EPs \cite{jia2022topological}. Another significant direction is to investigate the role of the higher-order EPs, especially the odd-order ones with anomalous behaviour, in designing non-Hermitian topological sensors \cite{PhysRevLett.125.180403}. 
Due to higher-order singular behaviour near a regularly behaved higher-order EP, the sensors based on such EPs are expected to show greater sensitivity than an EP$_2$, and the existence of mixed-type EPs may enable us to tune the sensitivity by tuning the parameter space \cite{PhysRevA.106.023508}.

\acknowledgments{
We acknowledge helpful discussions with Emil~J.~Bergholtz, Jan Budich, Lukas K\"onig, Marcus St\r alhammar, and Zhi Li. K.Y. is supported by the Swedish Research Council (VR, grant 2018-00313) and the Wallenberg Academy Fellows program (2018.0460) Fellows program of the Knut and Alice Wallenberg Foundation, and ANR-DFG project (TWISTGRAPH).
}

\bibliography{ref}

\appendix

\section{Exceptional degeneracy under sublattice symmetry}
\label{secsupp1}

When a symmetry is imposed, the standard method for obtaining the EP parameter space (see Appendix~\ref{secsupp3}) can be very complicated to employ in practice. Therefore, we adopt a more direct way to find the EP parameter space under sublattice symmetry, which employs the algebraic connections between $\mathbf B$ and $\mathbf B'$ as linear transformation operators. In this appendix, we demonstrate this method for the $N=2$ case, where we can obtain closed-form expressions. We use $\mathbb C^\times$ to represent the set of all complex numbers $z\ne 0$. We also introduce the notation $\mathcal{J}_n$ to denote the set of nondegenerate matrices commuting with the Jordan block $J_n$. In fact, $\mathcal{J}_n$ is given by all upper-triangular translational-invariant matrices \cite{PhysRevA.102.032216} \footnote{Note that our $\mathcal{J}_n$ corresponds to $\mathbb C^\times\times\mathcal{J}_n$ in Ref.~\cite{PhysRevA.102.032216} and our $\mathcal{EP}_n$ includes the $n^{\text{th}}$ order EP of all energy spectra.}.

To get the $SU(2)$-invariant doublet of EP$_2$'s, the Hamiltonian is determined by $\mathbf B$ or $\mathbf B'$ with a second-order EP. Hence, the corresponding parameter space $\mathcal{EP}_2$ is given by $ GL(2)/ \mathcal J_2 $.

For $\mathcal{EP}_4$, we notice that $\dim(\ker \mathbf B)+\dim(\ker \mathbf B')=1$, according to the discussions in the main text. Assuming that $\dim(\ker \mathbf B)=1,\dim(\ker \mathbf B')=0$, without any loss of generality, $\mathbf B'$ is invertible. As we have shown in Sec.~\ref{sc_oddEP}, the matrix $ \mathbf B  \cdot  \mathbf B'  $ must be similar to $J_2(0)$, which means $ \mathbf B   \cdot  \mathbf B'  \cdot   \mathbf B   \cdot  \mathbf B'  = \mathbf B  \cdot  \mathbf B'  \cdot \mathbf B=0$. 
There can be two scenarios according to whether $\mathbf B$ is diagonalizable or non-diagonalizable:
\begin{enumerate}
\item
When $\mathbf B$ is diagonalizable, let the eigenvectors of $\mathbf B$ be $\chi_1$ and $\chi_2$. We choose $\chi_1\in \ker \mathbf B$. In order to have $ \mathbf B  \cdot \mathbf B'   \cdot  \mathbf B  =0$, it is enough to have $ \left( \mathbf B   \cdot  \mathbf B'  \cdot  \mathbf B  \right)   \chi_2=0$. Since $\chi_2$ is an eigenvector with a nonzero eigenvalue, this is equivalent to $ \left( \mathbf B   \cdot  \mathbf B'   \right)  \chi_2=0$, implying that $\mathbf B' \, \chi_2\in\ker\mathbf B$. Switching to the basis formed by $\chi_1$ and $\chi_2$, we get
\begin{align}
   \mathbf B'=\begin{pmatrix}
b'_1 & b'_2\\
b'_3 &0 \\
\end{pmatrix}
\textrm{ when }  \mathbf B=\begin{pmatrix}
0 & 0\\
0 &b_4 \\
\end{pmatrix},
\end{align}
with $b_4$ denoting the eigenvalue corresponding to $\chi_2$. In order to ensure that $\mathbf B'$ invertible, we need $b'_2 \,b'_3\ne 0$. 

\item
When $\mathbf B$ is not diagonalizable, it is equal to $J_2(0)$ in a basis formed by two linearly independent vectors $\chi_1$ and $\chi_2$, still with $\chi_1\in \ker \mathbf B$. Here also, we only need to have $\left( \mathbf B \cdot \mathbf B' \cdot \mathbf B  \right) \chi_2=0$, which is now equivalent to $\left( \mathbf B  \cdot  \mathbf B' \right)\chi_1=0$. This tells us that $\mathbf B' \,\chi_1\propto\chi_1$, i.e., $\chi_1$ is also an eigenvector of $\mathbf B'$. Switching to the basis formed by $\chi_1$ and $\chi_2$, we get
\begin{align}
    \mathbf B'=\begin{pmatrix}
b'_1 & b'_2\\
0 & b'_4 \\
\end{pmatrix} 
\text{ when } 
\mathbf B=\begin{pmatrix}
0 & 1\\
0 &0 \\
\end{pmatrix}.
\end{align}
The invertibility of $\mathbf B'$ requires that $b'_1 \, b'_4\ne 0$. Therefore, we find that the parameter space
$\mathcal{EP}_4$ comprises two sets: $\mathbb Z_2\times\mathbb C\times (\mathbb C^\times)^3\times GL(2)/(\mathbb C^\times)^2$ and
$ \mathbb Z_2 \times \mathbb C\times (\mathbb C^\times)^2\times GL(2)/\mathcal J_2$. The $\mathbb Z_2$ part in either set comes from the symmetry under $\mathbf B \leftrightarrow \mathbf B'$. 
\end{enumerate}

The space $\mathcal{EP}_3$, as shown in Sec.~\ref{sc_oddEP}, is restricted to obey $\dim(\ker \mathbf B)=\dim(\ker \mathbf B')=1$. Basic linear algebra then tells us that their corresponding image dimensions are also equal to one, i.e., $\dim(\textrm{im} \, \mathbf B)=\dim(\textrm{im} \, \mathbf B')=1$. Let the corresponding eigenvectors be
$\chi_1$ and $\psi_1$, such that $\mathbf B \, \chi_1=0 $ and $\mathbf B' \, \psi_1=0$. It is straightforward to verify that $(0, \chi_1^T)^T$ and $(\psi_1^T, 0)^T$ are eigenvectors of $H$. We assume that $(0,\,\chi_1^T)^T$ belongs to a generalized eigenspace of dimension three. Hence, there exists a vector $(\psi_2,\chi_2)$ such that $H \,(\psi_2^T, \chi_2^T)^T=(0,\chi_1^T)^T$. This implies $\chi_2\in \ker \mathbf B$ and $-i \, \mathbf B'\psi_2 =\chi_1 $, requiring $\textrm{im} \,\mathbf B' =\ker \mathbf B$. We can choose $\psi_2$ to be in the subspace complementary to that of $\psi_1$ (i.e., $\psi_2\in \{\mathbb C^2-(\ker \mathbf B')\}$) and set $\chi_2=0$. In order to form a three-dimensional generalized eigenspace, we need a third linearly-independent vector $(\psi_3, \chi_3)$, such that $H \,(\psi_3^T,  \chi_3^T)^T=(\psi_2^T , 0)^T$. From this relation, we have $\psi_3\in \ker\mathbf B'$ and $\textrm{im} \,\mathbf B\ne\ker \mathbf B'$, which enforces the condition $\psi_2\in \textrm{im} \,\mathbf B$ -- therefore we can choose $\psi_3=0$ and $\chi_3\in(\ker\mathbf B)_\perp$. One can verify that the four vectors -- $(0,\,\chi_1^T)^T$, $(\psi_1^T, 0)^T$, $(\psi_2^T, 0)^T$, and $(0,\chi_3^T)^T$ -- that we have just constructed, are linearly-independent. To summarize, once the matrix $\mathbf B$ is fixed, the image of $\mathbf B'$ also gets fixed, and $\ker \mathbf B'$ must be different from $\textrm{im} \,\mathbf B$.
Since $\dim(\ker\mathbf B')=1$, the matrix $\mathbf B'$ is determined by $\mathbf B$ up to a nonzero vector (characterizing the ratio between the first and second columns of $\mathbf B'$). The matrix $\mathbf B$ can be built from two linearly dependent row vectors:
\begin{align}
  \mathbf B=\begin{pmatrix}
   p_1\, u_1 &p_1 \,u_2\\
   p_2 \,u_1 & p_2 \,u_2
\end{pmatrix},
\end{align}
because its kernel is one-dimensional.
Here at least one of $p_1$ and $p_2$ is nonzero and so are $u_1,u_2$. The kernel of $\mathbf B$ is generated by the vector $(u_2,\,-u_1)^T$, and its image is generated by $(p_1,\,p_2)^T$. According to the relations between $\mathbf B$ and $\mathbf B'$, we have
\begin{align}
    \mathbf B'=\begin{pmatrix}
  p'_2 \, u_2  & -p'_1 \, u_2 \\
  -p'_2 \, u_1  &  p'_1 \,u_1 
\end{pmatrix},
\end{align}
where $(p'_1,p'_2)$ is not collinear with $(p_1,p_2)$. We observe that all the pairs $(u_1,\,u_2)$, $(p_1,\,p_2)$, and $(p'_1,\,p'_2)$ exclude the origin $(0,0)$. Since $\mathbf B$ is invariant under the transformations $u_i\to z \,u_i$,
$p_i\to p'_i/z$, and $p'_i\to p'_i/z$, its parameter space is represented by $\mathbb C^{2\times}\times\mathbb C^{2\times}\times (\mathbb C^{2}-\mathbb C)/\mathbb C^\times$. This leads to the final result that $\mathcal{EP}_3$ is given by $\mathbb Z_2\times\mathbb C^{2\times}\times\mathbb C^{2\times}\times (\mathbb C^{2}-\mathbb C)/\mathbb C^\times$.

\section{Solutions for eigenvectors near an EP}
\label{secsupp2}

In this appendix, we work out the explicit expressions for the eigenvalues and eigenvectors near the EP$_4$ and EP$_3$ studied in Sec.~\ref{sc_eq2d}. Near the EP$_4$, the off-diagonal submatrices of the Hamiltonian take the forms:
\begin{align}
    \mathbf B(\mathbf q_\ast+\delta \mathbf q)&\simeq\begin{pmatrix}
 v_1(\theta)  \, |\delta\mathbf q|  & b_2 \\
0 & v_4(\theta)  \, |\delta\mathbf q|  \\
\end{pmatrix},
\nonumber\\
\mathbf B'(\mathbf q_\ast+\delta \mathbf q)&\simeq
\begin{pmatrix}
 b'_1  & 0 \\
  v'_3(\theta)  \, |\delta\mathbf q| & b'_4 \\
\end{pmatrix},
\end{align}
to leading order in the powers of $|\delta \mathbf q|$.
Their product matrix is given by
\begin{align}
    &\mathbf B'(\mathbf q_\ast+\delta \mathbf q)\cdot
    \mathbf B(\mathbf q_\ast+\delta \mathbf q)\nonumber\\
    \simeq &
    \begin{pmatrix}
 b'_1 \,v_1(\theta)  \, |\delta\mathbf q|  & b_2\,b'_1 \\
v_1(\theta)\,v'_3(\theta)  \, |\delta\mathbf q|^2  &
\left  [b_2 \,v'_3(\theta)+v_4(\theta)\,b'_4 \right ] |\delta\mathbf q|  \\
\end{pmatrix},
\end{align}
\begin{widetext}
with eigenvalues
\begin{align}
    \lambda_{a}=\frac{1}{2}
    \left[b'_1\,v_1+b_2 \,v'_3+b'_4\,v_4 +(-1)^{a+1}
    \sqrt{(b'_1\,v_1+b_2 \,v'_3+b'_4 \,v_4)^2-4\,b'_4\,v_4\,b'_1\,v_1}\right]|\delta\mathbf q|\,,
\text{ with } a \in \lbrace 1,2 \rbrace.    
\end{align}
The four eigenvalues $E$ of the Hamiltonian are therefore given by $\pm\sqrt{\lambda_1}$ and $\pm\sqrt{\lambda_2}$. The (unnormalized) eigenvectors of $\mathbf B'\cdot \mathbf B$ are
\begin{align}
    \chi^T_{a}=\left(1,-\frac{2 \,v_1 \,v'_3 \,|\delta\mathbf q|}
    {b_2 \,v'_3+b'_4 \,v_4-b'_1 \,v_1 +(-1)^a
 \sqrt{(b_2 \,v'_3+b'_4 \,v_4-b'_1 \,v_1)^2
 +4 \,b'_1 \,v_1 \, b_2 \, v'_3}}\right)\simeq
 \left (1,\,{\mathcal O} (|\delta\mathbf q|) \right ),
\end{align}
and hence are seen to converge to $(1,0)$ at the EP.
Using the relation $\psi_a = i\, \mathbf B\, \chi_a / E $ for $E \neq 0$, we deduce that $\psi^T_{a}\simeq
\bigg ({\mathcal O} \big (|\delta\mathbf q|^{1/2}\big ), \,{\mathcal O} \big(|\delta\mathbf q|^{3/2}\big) \bigg)$, giving the four eigenvectors of the Hamiltonian as $(\pm\psi_a^T ,\chi_a^T )^T $. Clearly, these four vectors collapse to $e_1= (0,0,1,0)^T$, as described in the main text.
\end{widetext}

As for the EP$_3$, since the exact expression is quite complicated, we only show the leading order terms. For {\it Path~1}, where all deviations from the EP are linear, we have
\begin{align}
      \mathbf B(\mathbf q_\ast+\delta \mathbf q)&\simeq\begin{pmatrix}
 v_1(0)  \, |\delta q_x|  & b_2 \\
v_3(0)  \, |\delta q_x|  & v_4(0)  \, |\delta q_x|  \\
\end{pmatrix},\nonumber\\
\mathbf B'(\mathbf q_\ast+\delta \mathbf q)&\simeq\begin{pmatrix}
 b'_1  & b'_2 \\
v'_3(0)  \, |\delta q_x|  & v'_4(0)  \, |\delta q_x|  \\
\end{pmatrix},
\end{align}
leading to
\begin{align}
    &\mathbf B'(\mathbf q_\ast+\delta \mathbf q)\cdot\mathbf B(\mathbf q_\ast+\delta \mathbf q)\nonumber\\
    \simeq&\begin{pmatrix}
 [b'_1v_1(0)+b'_2v_3(0)] \, |\delta q_x|  & b_2 \, b'_1 \\
[v_1(0) \, v'_3(0)+v_3(0)\, v'_4(0)] \, |\delta q_x|^2  & b_2 \, v'_3(0)  \, |\delta q_x| \\
\end{pmatrix}\nonumber\\
=&p_2\begin{pmatrix}
 p_1\, |\delta q_x|  & 1 \\
p_3 \, |\delta q_x|^2  & p_4 \, |\delta q_x|
\end{pmatrix}.
\end{align}
Since the eigenvalues of the product matrix are
\begin{align}
    \lambda_{a}\simeq&
    \frac{p_2}{2}\left [ p_1+p_4  + (-1)^a \,
  \sqrt{(p_1-p_2)^2+4 \,p_3}\right] |\delta q_x|\nonumber
 \\&+{\mathcal O} (|\delta q_x|^2)
 \text{ with } a \in \lbrace 1, 2\rbrace  \,,
\end{align}
the eigenvalues of the Hamiltonian are of ${\mathcal O} (\sqrt{|\delta q_x|})$. The corresponding eigenvectors are given by
\begin{align}
    \chi_{a} \simeq& \left( 1,\,\frac{2\,p_3 \,|\delta q_x|}
  {p_1-p_4 + (-1)^a \, \sqrt{(p_1-p_2)^2+4 \,p_3}}\right)^T\nonumber\\
  \simeq &
 \Big   (1,\,{\mathcal O} (|\delta q_x|) \Big )^T .
\end{align}
According to the relation $\psi_a =i\, \mathbf B\, \chi_a / E$ [with $i \,\mathbf B \,\chi_a \simeq
\Big ({\mathcal O} (|\delta q_x|),\,{\mathcal O} (|\delta q_x|) \Big )^T$], each $\psi_{a}$ vanishes as $|\delta q_x|\to 0$. Overall, the four eigenvectors $(\pm \psi_a^T, \chi_a^T)^T $ are seen to collapse to $e_1 = (0,\,0, \,1, \,0)^T$, resulting in the EP$_3$ behaving as a typical EP$_4$, as far as the eigenvector-coalescence is concerned.

When we consider {\it Path~2} for approaching the EP$_3$, the off-diagonal matrices are given by
\begin{align}
   &\mathbf B(\mathbf q_\ast+\delta \mathbf q)\simeq
      \begin{pmatrix}
 v_1 \,|\delta q_y|  & b_2 \\
v_3 \,|\delta q_y|  & v_4 \,|\delta q_y|  \\
\end{pmatrix} \nonumber\\ \text{ and }
&\mathbf B'(\mathbf q_\ast+\delta \mathbf q)\simeq
\begin{pmatrix}
 b'_1  & b'_2 \\
r'_3 \,|\delta q_y|^2  & r'_4 \,|\delta q_y|^2  \\
\end{pmatrix},
\end{align}
leading to
\begin{align}
     &\mathbf B'(\mathbf q_\ast+\delta \mathbf q)\cdot
    \mathbf B(\mathbf q_\ast+\delta \mathbf q)\nonumber\\
    \simeq &\begin{pmatrix}
 (b'_1 \,v_1+b'_2 \,v_3) \,|\delta q_y|  & b_2 \,b'_1 \\
(v_1\, r'_3+v_3\,r'_4)\,|\delta q_y|^3  & b_2 \,r'_3  \, |\delta q_y|^2  \\
\end{pmatrix}\nonumber \\=&p'_2
\begin{pmatrix}
 p'_1 \,|\delta q_y|  & 1 \\
p'_3\, |\delta q_y|^3  & p'_4 \,|\delta q_y|^2
\end{pmatrix}.
\end{align}
Unlike the results for {\it Path 1}, the two eigenvalues of the above matrix are given by
\begin{align}
   \lambda_1 \simeq p'_2 \,p'_1\, |\delta q_y| \text{ and }
    \lambda_2 \simeq\frac{p'_2}{p'_1}\left(p_1'\,p'_4-p'_3\right)|\delta q_y|^2\,,
\end{align}
with their corresponding eigenvectors
\begin{align}
    \chi_1 =\left(1, \,\frac{p'_3 \,|\delta q_y|^2}{p'_1}\right)^T
\text{ and }
\chi_2 =\left(1, \,-\frac{2\,p'_1 \,p'_3 \,|\delta q_y|}
    {2\,p'_4 \,p'_1+2 \,p'_3}\right)^T
\end{align}
showing distinct scalings. Noting that $\psi_1 \simeq \left ({\mathcal O} ( \sqrt {|\delta q_y|}), \,  {\mathcal O} ( \sqrt {|\delta q_y|} \right )^T$ and $\psi_2  \simeq \big ({\mathcal O} (1), \, {\mathcal O}(1) \big )^T$, the eigenvectors $(\pm \psi_1^T ,\chi_1^T)^T$ go to $e_1$ at the EP, while $(\pm \psi_2^T ,\chi_2^T)^T$ do not collapse to any of the eigenvectors $e_1$ and $e_2$ of the EP$_3$.

\section{Exceptional degeneracy in the absence of sublattice symmetry}
\label{secsupp3}

In order to figure out the eigenspace of an $n\times n$ matrix $ {\mathcal D} $, it boils down to finding a nondegenerate matrix $V\in GL_n(\mathbb C)$, such that $V \, {\mathcal D} \,V^{-1}$ is equal to a block diagonal matrix $M_d=\mathrm{diag}\{J_{i_1}(E_{1}), \,J_{i_2}(E_{2}), \, \dots\}$. All information about exceptional degeneracy is encoded in $M_d$. Let us denote the matrices commuting with $M_d$ as $S_d$, which may also be called the stabilizer of $M_d$ under the action of $GL_n(\mathbb C)$. The possible distinct matrices sharing the same exceptional structure are then given by the orbit $GL_n(\mathbb C)/S_d$. Thus, for a given $M_d$, $GL_n(\mathbb C)/S_d$ is the parameter space of the EP at the energy $(E_1,E_2\dots)$.

Let us now demonstrate how the parameter space of an EP looks like by focussing on the case of $n=4$. All $4\times 4$ complex matrices form a $16$-dimensional complex space $M_4(\mathbb C)=\mathbb C^{16}$. The parameter space of an EP is thus a (topological) subspace of this $\mathbb C^{16}$ and, compared to Hermitian degeneracies, the space of an exceptional degeneracy has a much richer structure. The constructions for the various possible cases are shown below:
\begin{enumerate}

\item
We first consider the scenario when all eigenvalues are degenerate, which consists of the highest-order EP, with the corresponding parameter space denoted as $\mathcal{EP}_4$ \cite{PhysRevA.102.032216}. Using the notations introduced in Appendix~\ref{secsupp1}, the Jordan block for the exceptional degeneracy is given by $J_4(E)$, and the $\mathcal EP_4$ is described by $\mathbb C\times GL_4(\mathbb C)/ \mathcal J_4$ [where the first $\mathbb C$ corresponds to the complex eigenvalue $E$ of $J_4(E)$]; its complex dimension is $4^2+1-4=13$.
The stabilizer $ \mathcal J_4$ is composed of polynomials of $J_n(0)$, with the condition that the coefficient of $\mathbb I_4$ is nonzero. The space $\mathcal{EP}_4$ is not simply connected, and is homotopically equivalent to $SU(4)/\mathbb Z_n$, where $\mathbb Z_4$ is the cyclic group formed by all fourth-order roots of unity \cite{PhysRevA.102.032216} -- this implies that $\mathcal{EP}_4$ has a nontrivial topology. A major difference from the degeneracies of Hermitian matrices stems from the fact that the transformation group $GL_4(\mathbb C)$, unlike the unitary group, is neither a closed subspace of $\mathbb C^{16}$ [it is an open subspace as the pre-image of $\det(M_4)\ne 0$], nor compact. Additionally, the parameter space of an EP at a given energy is not closed, as we have already shown in the main text. This is in sharp contrast with the parameter space of highest-order Hermitian degeneracy. The latter is given by $\mathbb C$, which is contractible, simply-connected, and closed in $\mathbb C^{16}$. It is described by matrices of the from $E\times \mathbb I_4$. The degeneracy parameter space at a given energy is simply a point. 

\item
An EP$_3$ is of intermediate order, and the space $\mathcal {EP}_3$ in $\mathbb C^{16}$ is represented by $V\textrm{diag}\{J_3(E_1), \,E_2\} \,V^{-1}$, with $V\in GL_4(\mathbb C)$. The parameters $E_1$ and $E_2$ form the space $\mathbb C^2$. In order to work out $\mathcal EP_3$, we need to quotient out those $V$ commuting with $\textrm{diag}\{J_3(E_1),E_2\}$. To do so, first we rewrite $\textrm{diag}\{J_3(E_1), \,E_2\}$ as $E_1 \,\mathbb I_4+\textrm{diag}\{J_3(0), \, E_2-E_1\}$. Since $\mathbb I_4$ commutes with any matrix, the problem is now reduced to finding the matrices commuting with $\textrm{diag}\{J_3(0), \, E_2-E_1\}$, which we denote as $\bar S$. The block form of $\bar S$ should satisfy 
\begin{align}
    &\bar S=\begin{pmatrix}
\mathbf S_1 &\mathbf S_2 \\
\mathbf S_3 &\mathbf S_4 \\
\end{pmatrix},\quad 
\mathbf S_1 \,J_3(0)=J_3(0)\,\mathbf S_1,\nonumber \\ &J_3(0)
\mathbf S_2=(E_2-E_1) \,\mathbf S_2\,, \, 
\mathbf S_3 \,J_3(0)=(E_2-E_1)\,\mathbf S_3\,,
\end{align}
with $\mathbf S_1$ representing a $3\times 3$ matrix and $\mathbf S_4$ denoting a complex number. When $E_1\ne E_2$, we must have $\mathbf S_2=0$ and $\mathbf S_3=0$. For $E_1=E_2$, they can be nonvanishing. The results are summarized as
\begin{align}
    \text{if } E_1\ne E_2 \,, \quad\bar S&=\textrm{diag}\left \lbrace
\sum_{m=0}^{m=2} s_1^{(m)} J^m_3(0),s_4 
\right \rbrace,\nonumber \\ &s^{(0)}_1s_4\ne 0;\nonumber\\
 \text{if } E_1=E_2\,, \quad \mathbf S_1&=\sum_{m=0}^{m=2} s_1^{(m)}\, J^m_3(0)\,,
\quad  \mathbf S_2=(s_2,0,0)^T,\nonumber \\ \mathbf S_3&=(0,0,s_3)\,,
\quad  \mathbf S_4=s_4\,,\quad  s^{(0)}_1 \,s_4\ne 0\,.
\end{align}
Thus, $\mathcal EP_3=U_1\cup U_2$, where $U_1$ and $U_2$ are two disjoint sets, with complex dimensions $14$ and $11$, respectively. The space $U_1$ consists of all matrices with $E_1 \neq  E_2$, i.e., $U_1=\mathrm{Conf}_2(\mathbb C)\times GL_4(\mathbb C)/(\mathbb C^\times\times\mathcal J_3)$ [where $\mathrm{Conf}_2(\mathbb C)$ is the second configuration space comprising all pairs $\lbrace E_1\in \mathbb C,E_2\in \mathbb C \rbrace $ with $E_1\ne E_2$]. $U_1$ characterizes all regular EP$_3$'s, where they exhibit the typical eigenvector-coalescence features, since there is a gap between the Jordan block and other levels. On the other hand, the space $U_2$ accounts for the case $E_1=E_2$ in the set $\lbrace E_1,E_2 \rbrace $, and is given by $\mathbb C\times GL_4(\mathbb C)/[\mathbb C^2\times \mathbb C^\times\times \mathcal J_3]$.

\item
The remaining exceptional degeneracy relevant to our discussions is EP$_2$. The space $\mathcal EP_2$ also contains those EPs that are of a mixed nature. But for the sake of simplicity, we neglect them, focussing only on regular EP$_2$'s. In this case, the Hamiltonian matrix takes the form $\mathrm{diag}[J_2(E_1),E_2,E_3]$, with distinct eigenvalues $E_1$, $E_2$, and $E_3$. The corresponding stabilizer turns out to be $\mathrm{diag}[\mathbf S_1,s_2,s_3]$, with $\mathbf S_1\in \mathcal{J}_2$. As a result, the regular part of $\mathcal EP_2$ is given by $GL_4(\mathbb C)\times \mathrm{Conf}_3(\mathbb C)/[(\mathbb C^\times)^2\times \mathcal J_2\times \mathbb Z_2]$, which is of complex dimension $15$, where the last $\mathbb Z_2$ comes from the general linear transformations that merely exchange $E_2$ and $E_3$. 

\end{enumerate}

 
\section{EP$_4$ with quartic-root singularity around it}
\label{sec_suppEP$_4$}

The EP$_4$ example provided in the main text has a square-root dispersion near the degeneracy. Here, we provide an example of a different EP$_4$which features a branch cut with quartic-root singularity.

As shown in Table.~\ref{tb_EPSpace}, the requirement for the existence of an EP$_4$ is to have $\mathbf B\cdot\mathbf B'$ proportional to a $2\times 2$ Jordan block, with at least one of the individual matrices (i.e., $\mathbf B$ or $\mathbf B'$) being non-invertible. To get a fourth-order root for the dispersion of the Hamiltonian, the eigenvalues of  $\mathbf B\cdot\mathbf B'$ should have a square-root dispersion. The typical form of $\mathbf B\cdot\mathbf B'$ then needs to be a Jordan matrix with a linear term $\sim|\delta \mathbf q|$ for the lower-left component. According to this logic, we can consider the forms:
\begin{align}
    \mathbf B(\mathbf q_\ast+\delta \mathbf q)&
    \simeq\begin{pmatrix}
0  & b_2 \\
v_3(\theta) \,|\delta \mathbf q| & 0  \\
\end{pmatrix},
\nonumber \\  
\mathbf B'(\mathbf q_\ast+\delta \mathbf q)&\simeq\begin{pmatrix}
 b'_1  & 0 \\
 0 & b'_4 \\
\end{pmatrix}.
\end{align}
In each position where the matrix element is put to zero, we have neglected possible $\mathcal{O} \big(|\delta\mathbf q| \big)$ terms, as they give a higher-order dispersion as explained in the main text. The product matrix is then given by
\begin{align}
    \mathbf B'(\mathbf q_\ast+\delta \mathbf q)\cdot
    \mathbf B(\mathbf q_\ast+\delta \mathbf q)\simeq \begin{pmatrix}
0  & b_2 \, b'_4 \\
 b'_1\, v_3(\theta) \,|\delta\mathbf q| & 0 \\
\end{pmatrix}.
\end{align}
The leading order expansion for an eigenvalue $\lambda$ of $\mathbf B\cdot\mathbf B'$ goes as
$\lambda\simeq \sqrt{b_2 \, b'_1 \, b'_4 \,v_3(\theta )
\, |\delta\mathbf q|}$. As the energy goes as $\sqrt{\lambda}$, we obtain a quartic-root behaviour in the vicinity of the EP$_4$.

\section{Lattice realizations for $N=2$ through the Yao-Lee model}
\label{secsupp4}

An example of $N=3$ flavours of fermions with sublattice symmetry is provided by the $SU(2)$ spin liquid model by Yao and Lee \cite{yao-lee}. We use two of its flavours to realize the exceptional points discussed in the main text. The Hamiltonian in this decorated honeycomb lattice [cf. Fig.~\ref{fighoney} (a)] is given by
\begin{align}
\hat H_{YL} =& J\sum_i \mathbf{S}^2_i + \sum_{\lambda\textrm{-link}~\avg{ij}} J_\lambda
\left( \tau_i^\lambda \,\tau_j^\lambda\right )
\left ( \mathbf S_i\cdot \mathbf S_j\right )\nonumber\\
&\text{ with } \lambda \in \lbrace 1,2,3 \rbrace ,
\nn \tau^1_i  =&1/2  + 2 \,\boldsymbol \sigma_{i,1}\cdot \boldsymbol \sigma_{i,2}\,,
\nonumber\\
\tau^2_i=&2\left ( \boldsymbol \sigma_{i,1}\cdot \boldsymbol \sigma_{i,3}
- \boldsymbol \sigma_{i,2}\cdot \boldsymbol \sigma_{i,3}
\right )/\sqrt{3}\,, \nonumber\\
\tau^3_i=&4 \, \boldsymbol \sigma_{i,1}\cdot 
\left (\mathbf S_{i,2}\times \mathbf S_{i,3} \right )/\sqrt{3}\,,
\label{eqsu2}
\end{align}
where the indices $i$ and $j$ label the triangles, and $\boldsymbol \sigma_{i,\alpha}$ denotes the vector spin-1/2 operator at site $\alpha \in \lbrace 1, 2, 3 \rbrace $ of the $i^{\rm{th}}$ triangle.
Furthermore, $\mathbf S_i= \boldsymbol \sigma_{i,1}
+ \boldsymbol \sigma_{i,2}+ \boldsymbol \sigma_{i,3}$ is the total spin operator of the $i^{\rm{th}}$ triangle. 
The coupling constant $J$ is the strength of the intra-triangle spin-exchange, while $J_\lambda$ describes the inter-triangle couplings on the $\lambda$-type link. There are three different types of links, $x$-, $y$-, and $z-$ links, represented by red, green and blue ones in Fig.~\ref{fighoney} respectively.
Since $\left [\mathbf S^2_i, \mathbf S_j \right ]=0$ and
$\left [\mathbf S^2_i, \tau^{\lambda}_j \right ]=0$, the operator $\mathbf S_i^2$ commutes with the Hamiltonian for all $i$. Hence, the total spin of each triangle is a good quantum number, which we can use to subdivide the Hilbert space.

Just like the case of Kitaev's model on the honeycomb lattice \cite{kitaev}, we first introduce the Majorana fermion representations for the Pauli matrices $\sigma_{i,\alpha} $ and $\tau^\beta_i$ as follows:
\begin{align}
\sigma_{i,\alpha}\, \tau^\beta_i =i\,\eta^\alpha_i \,d^\beta_i\,,\quad 
\sigma_{\alpha,i} =
-\frac{i}{2}\, \epsilon^{\alpha \beta\gamma}\, \eta^\beta_i \,\eta^\gamma_i\,,\nonumber\\
\tau^\alpha_i =-\frac{i}{2} \,
\epsilon^{\alpha\beta\gamma}\,d^\beta_i d^\gamma_i\,,
\text{ with } \alpha,\beta \in \lbrace 1,2,3 \rbrace,
\end{align}
where  $ \eta^\alpha_i$ and $d^\alpha_i$ are Majorana fermion operators (i.e., ${ \eta^\alpha_i}^\dagger = \eta^\alpha_i$
and ${ d^\alpha_i}^\dagger = d^\alpha_i$). The Hilbert space is enlarged in the Majorana representation and the physical states are those invariant under a $\mathbb Z_2$ gauge transformation.
Using the above notation, we can reexpress the Hamiltonian $\hat H_{YL}$ as
\begin{align}
\hat H_{YL}& =Q\,\hat{\mathcal H}_{eYL} \,Q\,,\nonumber\\
\hat {\mathcal H}_{eYL}&=\sum_{\avg{ij}}
J_{ij} \, u_{ij} \left [i\,\eta^1_i \,\eta^1_j + i\, \eta^2_i \,\eta^2_j 
+ i\,\eta^3_i \,\eta^3_j \right ] ,
\label{eq:fer_Ham}
\end{align}
where $ u_{ij}=-i\,d^\lambda_i\, d^\lambda_j$, $J_{ij}=J_\lambda/4$ on the $\lambda$-type
link $\avg{ij}$, and $Q$ is the projection operator on the physical states.
Because $\left [ u_{ij}, \hat{\mathcal H} \right ]=0$ and $\left [ u_{ij}, u_{i'j'}\right ]=0$,
the eigenvalues (which take the values $\pm 1$) of the $ u_{ij}$'s are good quantum numbers.
From its form, it is clear that $\hat{\mathcal H}_{eYL}$ describes three flavours of Majorana fermions, coupled with the background $Z_2$ gauge fields denoted by $u_{ij}$.
One can verify that $\hat{\mathcal H}_{eYL} $ is invariant under the local $Z_2$ gauge transformation, which takes
$ \eta^\alpha_i\to \Lambda_i \, \eta^\alpha_i$ and $u_{ij}\to \Lambda_i \,u_{ij}\,\Lambda_j$,
with $\Lambda_i=\pm 1$.
In addition to the $Z_2$ gauge symmetry, the system has a global $SO(3)$ symmetry, which rotates among the three flavours of Majorana fermions, and is a consequence of the $SU(2)$ symmetry of the original spin model. 

Each Majorana flavour $c^\alpha$ has a Hamiltonian identical to the single Majorana flavour in Kitaev's honeycomb model \cite{kitaev}, and hence the Yao-Lee model effectively gives us three copies of the Kitaev model. The spectrum of the Majorana fermions is gapless, while the $Z_2$ gauge field has a finite gap from the flux-fee configuration given by $u_{ij}=1$. The low-energy theory of the $SU(2)$ model is thus captured by setting $ u_{ij}=1$, leading to the momentum-space Majorana Hamiltonian 
\begin{align}
\hat H_m&=c^T \, H_{m}\, c\,, \nonumber\\
c&= \begin{pmatrix}
 c_1^{1}(\mathbf q) & c_1^{2}(\mathbf q) & c_1^{3}(\mathbf q)
 & c_2^{1}(\mathbf q) & c_2^{2}(\mathbf q) & c_2^{3}(\mathbf q) \\
\end{pmatrix}^T ,
\end{align}
where
\begin{align}
\label{eqyao-lee}
H_{m} &=\begin{pmatrix}
0 & i\,\mathbf{A}(\mathbf{q}) \\
-i\,\mathbf{A}^T(- \mathbf{q}) & 0 \\
\end{pmatrix},\quad 
\mathbf{A}(\mathbf{q})
=\mathbb{I}_3 \otimes \tilde A(\mathbf q),\nonumber\\  
\tilde A(\mathbf q)&= 2\left( J_1\,e^{i\,\mathbf q\cdot\mathbf r_1}
 + J_2 \,e^{i\,\mathbf q\cdot\mathbf r_2} + J_3\right).
\end{align}
Here, $c^\alpha$ denotes the Fourier transform of a real-space $\eta^\alpha $-operator, and the subscripts $1$ and $2$ refer to the two sublattice sites $A$ and $B$ of the honeycomb lattice. Furthermore, the unit cell vectors of the triangular lattice, generating the honeycomb lattice, have been labelled by $\mathbf r_1$ and $\mathbf r_2$. For notational convenience, we also introduce a third vector defined by $\mathbf r_3=\mathbf r_1-\mathbf r_2$.

Before constructing lattice Hamiltonians harbouring higher-order EPs, let us first review the second-order EPs obtained in a non-Hermitian extension of the Kitaev model, studied in Ref.~\cite{kang-emil}.
The momentum-space Hamiltonian takes the form:
\begin{align}
\label{eqkit}
H_{K} =\begin{pmatrix}
0 & i\,\tilde  A(\mathbf{q}) \\
-i\,\tilde A(- \mathbf{q}) & 0\\
\end{pmatrix}, 
\end{align}
where the spin-spin coupling constants are tuned to complex values, parametrized as $J_1 =|J_1| \,\exp(i\,\phi_1) $ and $ J_2  = |J_2| \,\exp(i \,\phi_2)$, and $J_3 $ (with $\phi_1$, $\phi_2$, and $J_3 $ constrained to be real numbers).
The Dirac points of the Majorana fermion dispersion for $\phi_1 =\phi_2 =0 $ morph into EPs, as nonzero values of $\phi_1 $ and $\phi_2 $ are turned on \cite{kang-emil}, and are located at
\begin{align}
 &\tilde q_{1(2)} = \pm \cos^{-1} 
\left(\frac{| J_{ 2 (1)}|^2-|J_{ 1( 2)}|^2-|J_{3}|^2}
{2\,| J_{1(2)}|\, | J_{3}|} \right) 
- {\phi}_{ 1 (2)}\,,\nonumber \\
&|J_{1}| \sin( \tilde  q_1+ {\phi}_1) = - |J_{2}| \sin( \tilde q_2 + {\phi}_2)\,,
\end{align}
where $\tilde {\mathbf{q} }=(\tilde q_1 ,\tilde q_2)$ are the coordinates of the momentum vector in the reciprocal lattice space, in the basis of the reciprocal lattice vectors. The second equation fixes the signs in the first. 
The exceptional nature stems from complex $J_\alpha$'s due to the fact that $\tilde A^\ast(\mathbf q)\ne 
\tilde A(-\mathbf q)$. 
There are pairs of EPs connected by Fermi arcs, and are thus robust against perturbations.

The $SO(3)$-extension of Eq.~\eqref{eqkit}, as shown in Eq.~\eqref{eqyao-lee}, has six bands, and thus has the possibility to host higher-order EPs. To start with, we can tune the $J_\alpha$'s into complex numbers, as illustrated above. However, this results only in a triplet of EP$_2$'s, each arising from one flavour of the Majorana fermions. In order to obtain higher-order EPs, we need to break the $SO(3)$-symmetry by introducing couplings between the three flavours in various ways, and/or using different values of the $J_\lambda$'s for the three flavours. For instance, for nearest-neighbour couplings between different flavours, the relevant spin operators take the form: $\sigma^\alpha_i \dots
\tau^\beta_i  \dots  \sigma^\gamma_j  \dots  \tau^\lambda_j  \dots$, with $i$ and $j$ here denoting the indices of the nearest-neighbour sites. As a concrete example, the operator $i\,\exp(i\,\mathbf q\cdot\mathbf r_1)\,
c^\alpha(-\mathbf q)\,c^\beta(\mathbf q)$ (with $\alpha \neq \beta $) translates into $ \sigma_{\alpha,i}\,\tau^1_{i}\,\tau^1_{j}\,\sigma_{\beta,j}$.

In order to have a non-Hermitian behaviour, we choose $J_1= \tilde J\,\exp(i\,\phi)$, and $J_2 =J_3 = \tilde J$, where $\tilde J$ and $\phi$ are real parameters. The EPs are assumed to appear at $\mathbf q = \mathbf q_\ast$, as before. We introduce the functions $g(\mathbf q)=\exp(i\,\mathbf q\cdot\mathbf r_1+i\,\phi)
+\, \exp(\mathbf q_\ast\cdot\mathbf r_2)+1$, and $h(\mathbf q)
=\,\exp(i \,\mathbf q_\ast\cdot\mathbf r_1-i\,\phi)
+\, \exp(i\,\mathbf q\cdot\mathbf r_2)+1$. One can verify that $g(\mathbf q_\ast)=h(-\mathbf q_\ast)=0$. Since both of these represent nearest-neighbour hoppings, they can be constructed via the spin operators as described in the earlier paragraph.

To realize an EP$_4$, one way is to consider the form:
\begin{align}
   H_{m}&=\begin{pmatrix}
0 & i\,\mathbf{A}(\mathbf{q}) \\
-i\,\mathbf{A}^T(- \mathbf{q}) & 0 \\
\end{pmatrix},\nonumber\\
\mathbf A(\mathbf q)&=
\mathrm{diag} \left \lbrace \mathbf B(\mathbf q),\tilde A_0(\mathbf q) \right \rbrace,
\nonumber\\  \mathbf B(\mathbf q)&=\begin{pmatrix}
 \tilde A(\mathbf q)  &z_1 \, g(-\mathbf q)+ z_2\, h(\mathbf q)\\
0 & \tilde A'(\mathbf q) \\
\end{pmatrix},
\end{align}
which affects only the couplings among the operators $c_1^{1}(\mathbf q)$, $c_1^{2}(\mathbf q)$, $c_2^{1}
(\mathbf q)$, and $c_2^{2}(\mathbf q)$. 
Here, $z_1$ and $z_2$ are the coupling constants for the $h(\mathbf q)$ and $g(-\mathbf q)$ hoppings. For the flavour $\alpha =2$, we have used a different coupling $\tilde A'(\mathbf q)$, which is obtained by adding $\tilde A(\mathbf q)$ to $h(\mathbf q)$ or $g(-\mathbf q)$. Note that, in the low-energy Majorana fermion model, we have $\mathbf B'(-\mathbf q)=\mathbf B^T(\mathbf q)$ due to the particle-hole symmetry. The coupling $\tilde A_0 = 2\left( J_1^{(0)}\,e^{i\,\mathbf q\cdot\mathbf r_1}
 + J_2 ^{(0)} \,e^{i\,\mathbf q\cdot\mathbf r_2} + J_3^{(0)} \right)$ corresponds to the flavour $\alpha = 3$, and can be composed of a real set of values for the $J_\lambda^{(0)}$'s (as in the Hermitian case), as the $2\times 2$ block of this flavour does not take part in the exceptional physics corresponding to the $4\times 4 $ block that we are tying to construct.

In order to realize an EP$_3$, we need to make the couplings $c^{2}_1 \, c^{1}_2$ and $c^{2}_1 \,c^{2}_2$ anisotropic around the EP. This can be done by combining functions related by some type of crystal symmetry. Let us assume that the function $f(\mathbf q)$ vanishes linearly in $  \delta \mathbf q$ near $\mathbf q_\ast$. Then, we can find another function $f(q_x,2 {q_{\ast}}_y-q_y)$, which is the mirror reflection of $f(q_x,q_y)$ with respect to $\mathbf q_\ast$. Near $\mathbf q_\ast$, the combined function $f(q_x,q_y)+f(q_x,2 {q_{\ast}}_y-q_y)$ has a vanishing first-order derivative along the $q_y$-direction, while its leading order Taylor expansion along the $q_x$-direction is still linear, resulting in the desired anisotropy. Using these functions, we can now construct
the Hamiltonian of the Majoranas as 
\begin{align}
   H_{m}&=\begin{pmatrix}
0 & i\,\mathbf{A}(\mathbf{q}) \\
-i\,\mathbf{A}^T(- \mathbf{q}) & 0 \\
\end{pmatrix},\nonumber\\
\mathbf A(\mathbf q)&=
\mathrm{diag}\left \lbrace \mathbf B(\mathbf q),\tilde A_0(\mathbf q) \right \rbrace ,
\nonumber\\
\mathbf B(\mathbf q)&=
\begin{pmatrix}
 \tilde A(\mathbf q)  &f_1(\mathbf q)+f_1(q_x,2 {q_{\ast}}_y-q_y)\\
z'_1 \, g(\mathbf q)+ z'_2\, h(-\mathbf q) & \tilde A'(\mathbf q)+
\tilde A'(q_x,2 {q_{\ast}}_y-q_y) \\
\end{pmatrix},
\end{align}
where $f_1 = z_1 \, g(-\mathbf q)+ z_2\, h(\mathbf q)$, and $\mathbf B'(-\mathbf q)=\mathbf B^T(\mathbf q)$. The coupling $\tilde A_0$ can be constructed from real $J_\lambda^{(0)}$'s, similar to the EP$_4$ case. However, we immediately realize that the mirror-symmetric part of $\tilde A'(\mathbf q)$ [i.e., $\tilde A'(q_x,2 {q_{\ast}}_y-q_y)$], added to the original Hermitian spin Hamiltonian, is not perturbatively small. Hence, the above construction may create flux-excitations in the corresponding spin model (so that we are no longer in the zero flux state). Nevertheless, for a purely fermionic model, this construction will work without involving such issues.

\end{document}